\documentclass[aps,showkeys,physrev,reprint,superscriptaddress,twocolumn]{revtex4-2}

\usepackage{graphicx}
\usepackage{bm}
\usepackage{ulem}
\usepackage{xcolor}
\usepackage{amsmath}
\usepackage{amssymb}
\usepackage{bbold}
\usepackage{physics}
\usepackage[export]{adjustbox}
\usepackage{placeins}
\usepackage{enumitem}

\newcommand{\nt}{\notag}
\newcommand{\del}{\partial}

\makeatletter

\newcommand*{\underarrow}{\def\@underarrow{\relax}\@ifstar{\@@underarrow}{\def\@underarrow{\hidewidth}\@@underarrow}}
\newcommand*{\@@underarrow}[2][]{\underset{\@underarrow\substack{\uparrow\if\relax\detokenize{#1}\relax\else\\#1\fi}\@underarrow}{#2}}

\newcommand*{\overarrow}{\def\@overarrow{\relax}\@ifstar{\@@overarrow}{\def\@overarrow{\hidewidth}\@@overarrow}}
\newcommand*{\@@overarrow}[2][]{\overset{\@overarrow\substack{\if\relax\detokenize{#1}\relax\else#1\\\fi\downarrow}\@overarrow}{#2}}
\makeatother

\begin{document}

\title{Enhanced detection of circularly polarized photons with topological materials}

\author{Hamideh Sharifpour}
\email[]{avz6gh@virginia.edu}
\affiliation{Department of Electrical and Computer Engineering, University of Virginia, Charlottesville, VA, 22904, USA}

\author{Avik W. Ghosh}
\email[]{ag7rq@virginia.edu}
\affiliation{Department of Electrical and Computer Engineering, University of Virginia, Charlottesville, VA, 22904, USA}

\author{George J. de Coster }
\email[]{george.j.decoster.civ@army.mil}
\affiliation{DEVCOM Army Research Laboratory, 2800 Powder Mill Road, Adelphi, MD, 20738 USA}
\affiliation{MIT Institute for Soldier Nanotechnologies, 500 Technology Square, Cambridge, MA, 02139, USA}

\date{\today}

\begin{abstract}

Topological insulators (TI) are highly attractive platforms for next-generation optoelectronic and photonic devices. Spin-momentum locking of topological surface states enhances their nonlinear optical responses and sensitivities, especially to circularly polarized light. Until now, theoretical investigations of nonlinear responses in TIs have been limited to microscopic calculations on analytical continuum models, or leveraging density functional theory based Hamiltonians. In this work, we expand beyond these two approaches by employing a nonlinear Kubo formalism to calculate second-order nonlinear optical conductivity in a slab geometry using symmetry informed tight binding models that accurately reproduce the conduction, valence and topological surface bands in Bi$_2$Se$_3$.  Our methodology enables us to study the layer resolved contribution to injection currents coupled to the incident electric field. We demonstrate that our technique can reveal how device engineering modifies elements of the nonlinear optical response such as the circular {and linear} photogalvanic effects by breaking inversion and time-reversal symmetry. {In particular, magnetization-induced symmetry breaking
enables nonlinear conductivity tensor components (e.g., $\sigma_{xyz}$) that are normally forbidden by symmetry, thereby directly modifying the circular photogalvanic effect.}  We find, in line with experiments, that the photogalvanic current is sensitive to field effects, Fermi level energy, gate voltage and the energy of incident light. Our computed midwavelength infrared (mid-IR) responsivity  $R \approx 0.170~\mathrm{\mu A/W}$ is comparable to reported TI and intrinsic 2D-material photodetectors. We further simulate experimentally unexplored methods to modify the circular photogalvanic effect such as proximitizing a magnetic field to one of the TI surface materials, suggesting a mechanism for optoelectronic tuning.
\end{abstract}

\maketitle

\section{Introduction}

Quantum materials serve as a rich platform for exploring unconventional transport and optical properties \cite{orenstein2021topology,keimer2017physics,de2017quantized}. In particular, topological insulators (TIs) are characterized by an insulating bulk with robust metallic surface states protected by time-reversal symmetry, generating an extensive landscape to explore fundamental physical principles and engineer next-generation optoelectronic devices \cite{hasan2010colloquium,kuznetsov2022topological}. The surface states of TIs are topologically distinct, featuring spin-momentum locking and a Dirac-like energy spectrum, which enables exotic responses to external electromagnetic fields \cite{xia2009discovery}. 
Among known TIs, bismuth selenide (Bi$_2$Se$_3$) has emerged as a model system due to its relatively simple electronic structure and modest bulk band gap, which allows for the experimental isolation of its surface states \cite{wang20212d}. Other compounds such as Bi$_2$Te$_3$, Sb$_2$Te$_3$, and their doped or alloyed analogs contribute to the family of three-dimensional TIs, all together offering  unprecedented control over their electronic and optical properties \cite{hsieh2009observation}.

Linear and nonlinear optical processes describe fundamentally distinct regimes of light-matter interactions, both relevant to TIs, particularly at their boundaries \cite{boyd2008nonlinear}. In the linear optical regime, the material response is directly proportional to the applied electric field $\mathbf{E}$, expressed as: $\mathbf{P} = \varepsilon_0 \chi^{(1)} \mathbf{E}$, where $\varepsilon_0$ is the vacuum permittivity and $\chi^{(1)}$ is the linear susceptibility \cite{sheik2000third}. Conversely, nonlinear optical effects arise at high light intensities, where the response exceeds linearity, and higher-order terms contribute to polarization:
$\mathbf{P} = \varepsilon_0 \left( \chi^{(1)} \mathbf{E} + \chi^{(2)} \mathbf{E}^2 + \chi^{(3)} \mathbf{E}^3 + \cdots \right)$, where $\chi^{(2)}$ and $\chi^{(3)}$ are the second- and third-order nonlinear susceptibilities, respectively\cite{aversa1995nonlinear}. These nonlinear contributions enable phenomena such as SHG, third-harmonic generation (THG) and sum- and difference-frequency generation. For example, second-harmonic generation (SHG) involves two photons at frequency $\omega$ combining to produce a new photon at $2\omega$ \cite{Shen2002}. Similarly, the current density $\mathbf{J}$, linked to optical conductivity, expands as:
\begin{equation}
\mathbf{J} = \sigma^{(1)} \mathbf{E} + \sigma^{(2)} \mathbf{E}^2 + \sigma^{(3)} \mathbf{E}^3 + \cdots,
\end{equation}
where $\sigma^{(n)}$ denotes the $n$-th order optical conductivity \cite{morimoto2016topological}. Unlike linear responses, higher-order components of the current can be generated under intense or structured light excitation, facilitating advanced processes like harmonic generation, optical rectification, and photovoltaic currents beyond conventional limits \cite{sipe2000second}.

A major nonlinear effect in TIs is referred to as the photogalvanic effect (PGE), in which light generates a non-zero current with no applied bias
\cite{hsieh2011nonlinear,pan2017helicity,Braun2016,Plank2018,connelly2024emergence}.
This effect originates from the lack of inversion symmetry, which induces asymmetric optical transitions in the electronic band structure and gives rise to a directional nonlinear optical response \cite{belinicher1980photogalvanic}. There are two sub-classifications of PGE: the circular photogalvanic effect (CPGE), which is sensitive to the helicity of circularly polarized light, and the linear photogalvanic effect (LPGE),  driven by linearly polarized light in response to an anisotropic optical transition \cite{bel2003circular}.
Of the two effects, the CPGE serves not only as a probe of hidden symmetries and topological properties, but also provides a direct insight into the Berry curvature and spin textures in momentum space \cite{hosur2011circular,Junck2013}. The presence of spin-momentum locking in TIs naturally favors CPGE - an isolated up (down) spin electron preferentially couples to right (left) circularly polarized light. However, due to symmetric distribution of states in their $\vec{k}$-space over which the current integrates, a chiral photon cannot produce a net momentum flow in a centrosymmetric Dirac cone through a linear optical process. In addition, while DC current density breaks time-reversal symmetry, linear photoabsorption does not, which is why we need to look at nonlinear response to seek out signatures of chirality.

Importantly, when inversion symmetry is broken, either intrinsically via structural modifications or extrinsically through applied fields or interfaces, CPGE signals can be significantly enhanced, making TIs with broken symmetries particularly appealing candidates for realizing chiral photodetectors and polarization-sensitive devices \cite{huang2021optical}.

In this work, we attempt a complete description of nonlinear helical photocurrents in topological materials by explicitly calculating the injection current from the second-order optical conductivity tensor $\sigma^{(2)}_{\mu\alpha\beta}(\omega)$. Physically, this represents a population imbalance in momentum-space generated by the optical field \cite{Braun2016,leppenen2023linear}. Another second-order response is the shift current, which, unlike the injection current, is caused by a real-space displacement of the wavepacket center during interband optical transitions \cite{sturman2020ballistic,Osterhoudt2019}. 
We construct a theoretical framework to explain nonlinear optical conductivity in three-dimensional TIs based on a slab Hamiltonian. Unlike previous works employing continuum models within Fermi's golden rule and the Boltzmann equation, this method inherently includes bulk and surface states and thereby all interband transitions  \cite{pan2017helicity,Junck2013}. In this manner we can solve layer-resolved contributions to CPGE and LPGE. We analyze the effects of inversion-symmetry breaking—induced by gating, finite slab thickness, and their corresponding electric-field profiles—on the magnitude and character of the photocurrents. We find that the nonlinear photocurrent response in TIs is strongly dependent on the gate voltage, the Fermi level, the incident light frequency and magnetic proximity. Notably, we investigate the effect of magnetization on CPGE, by including in our model a  ferromagnetic layer deposited on the TI. We discover that breaking time-reversal and inversion symmetry by this top layer magnet provides an additional knob that can be tuned to boost, suppress, or even reverse helicity-dependent photocurrents, through a redistribution of Berry curvature from inside the magnetic gap to right above the edges.

Our treatment, which separates out bulk, surface and interlayer contributions, is in line with recent experiments shown in Fig.\ref{fig:6}, and provides an approach towards production of next-generation quantum and chiral-sensitive optoelectronic devices  working in the terahertz to mid-infrared regime.

\section{Theoretical Model and Methodology}
\subsection{Tight Binding Model}
We employ a lattice-regularized effective Hamiltonian to describe a three-dimensional TI \cite{mao2011tight,liu2010model,zhang2009topological}. The system is modeled as a slab geometry comprising atomic layers stacked along the $z$-direction. This has the desired effect of inherently containing the bulk and topological surface states so that the interband transitions from surface to bulk states which underlie CPGE and LPGE in Bi$_2$Se$_3$ are captured \textit{a priori}. The approach eliminates the need to explicitly solve for the bulk and surface wavefunctions in the continuum model or to  include each set of transitions manually \cite{Junck2013,pan2017helicity,Xie2025}. 

\begin{figure}[t!] 
  \centering
\includegraphics[width=1\columnwidth]{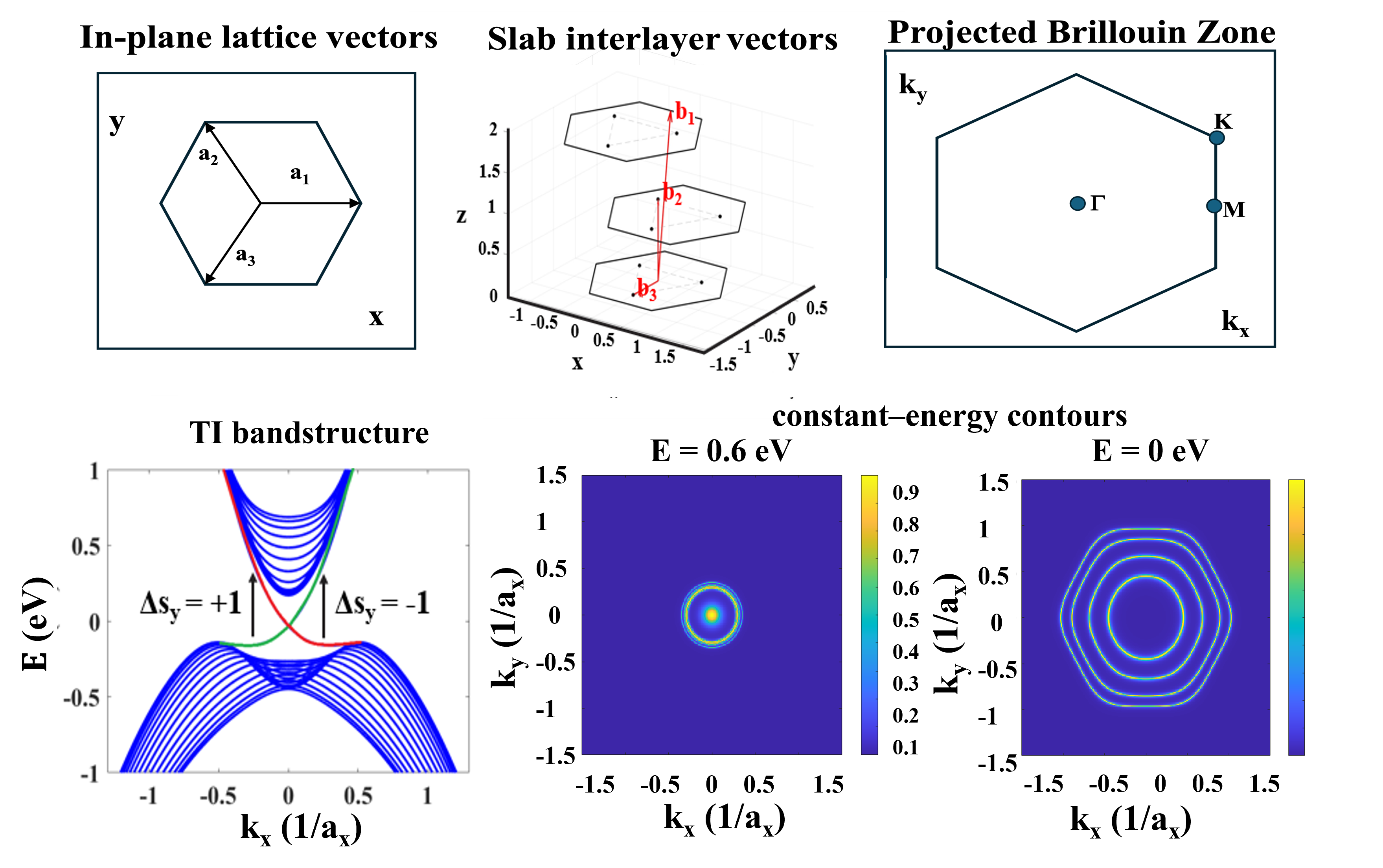}\\

 \caption{Top: In-plane lattice unit cell, inter-plane slab vectors, and 2D projection of first Brillouin zone for Bi$_2$Se$_3$. Bottom: Electronic structure of Bi$_2$Se$_3$: slab band structure showing a single Dirac cone bridging the bulk gap, red and green corresponding to $y$-spin polarization $\langle S_y \rangle = +1$ and $-1$ respectively. Band structure parameters are fitted to the conduction and valence bands of Bi$_2$Se$_3$; the constant–energy contours at $E=\pm0.6$ \text{eV} show that warping is only significant in the valence band but insignificant in the conduction band due to the small effective mass of the latter.}
  \label{fig:01}
\end{figure}

The low-energy Hamiltonian for in-plane momentum components $k_x$ and $k_y$ is expressed in a $4N_z \times 4N_z$ matrix form as:
\begin{eqnarray}
\mathcal{H} &=& \mathbf{c}^\dagger(\mathbf{k}) H_{\text{tot}} \mathbf{c}(\mathbf{k})~,\\
H_{\text{tot}} &=& H_0 \otimes \mathbb{1}_{N_z\times N_z} + H_z~,
\end{eqnarray}
where $H_0$ is the in-plane on-site Hamiltonian, $\otimes$ represents a tensor product, and $H_z$ encodes interlayer coupling terms. The ladder operators $c_{z,s}(\mathbf{k})$ are $4 N_z$ dimensional vectors Fourier transformed in the 2D-plane (i.e. $\mathbf{k} = (k_x, k_y)$). The Hamiltonian $H_{\text{tot}} $ is constructed using Dirac matrices derived from Pauli matrices $\sigma_i$. The momentum-dependent on-site Hamiltonian $H_0$ includes terms capturing lattice symmetries and spin-orbit coupling:
\begin{equation}
H_0(\mathbf{k}) = \varepsilon_0(\mathbf{k}) \mathbb{1}_{4\times4} + \sum_{i=1}^5 \varepsilon_i(\mathbf{k}) \Gamma_i,
\end{equation}
where $\varepsilon_i(\mathbf{k})$ are scalar functions encoding energy dispersions, and the $4 \times 4$ Dirac matrices $\Gamma_i$ are defined as:
\begin{equation}\label{eq:gamma-defs}
\begin{aligned}
\Gamma_1 &= \sigma_x \otimes \tau_x, &
\Gamma_2 &= \sigma_y \otimes \tau_x, &
\Gamma_3 &= \sigma_z \otimes \tau_x, \\
\Gamma_4 &= I_2 \otimes \tau_y, &
\Gamma_5 &= I_2 \otimes \tau_z .
\end{aligned}
\end{equation}
Here, $\sigma_{x,y,z}$ and $\tau_{x,y,z}$ are the Pauli matrices, and $I_2$ is the $2 \times 2$ identity matrix. The explicit energy terms are:
\begin{equation}
\label{eq:eps-k}
\begin{aligned}
\varepsilon_0(\mathbf{k}) &= 2A_0 \sum_{j=1}^{3} \cos\!\big(\mathbf{k}\!\cdot\!\mathbf{a}_j\big) - 6A_0, \\
\varepsilon_1(\mathbf{k}) &= -2A_{14} \sin\Omega  \Big[ \sin\!\big(\mathbf{k}\!\cdot\!\mathbf{a}_2\big) - \sin\!\big(\mathbf{k}\!\cdot\!\mathbf{a}_3\big) \Big], \\
\varepsilon_2(\mathbf{k}) &= -2A_{14} \Big[ \sin\!\big(\mathbf{k}\!\cdot\!\mathbf{a}_1\big) \nt\\
&~ ~~~ + \cos\Omega  \big( \sin\!\big(\mathbf{k}\!\cdot\!\mathbf{a}_2\big) + \sin\!\big(\mathbf{k}\!\cdot\!\mathbf{a}_3\big) \big) \Big], \\
\varepsilon_3(\mathbf{k}) &= 2A_{12} \sum_{j=1}^{3} \sin\!\big(\mathbf{k}\!\cdot\!\mathbf{a}_j\big), \\
\varepsilon_4(\mathbf{k}) &= 0, \\
\varepsilon_5(\mathbf{k}) &= 2A_{11} \sum_{j=1}^{3} \cos\!\big(\mathbf{k}\!\cdot\!\mathbf{a}_j\big) + m_{11}.
\end{aligned}
\end{equation}
The in-plane lattice vectors for the van der Waals planes 
(Fig.~\ref{fig:01}) are:
\begin{align}
\mathbf{a}_1 &= (a, 0, 0), 
~\mathbf{a}_2 = \left( -\tfrac{a}{2}, \tfrac{\sqrt{3}a}{2}, 0 \right), \notag \\
\mathbf{a}_3 &= \left( -\tfrac{a}{2}, -\tfrac{\sqrt{3}a}{2}, 0 \right) \label{eq:lattice-vectors}
\end{align}
and $\Omega = -\frac{2\pi}{3}$ is a geometric parameter representing the relative phase between hopping channels associated with the 120$^0$ angle-separated vectors $\vec{a}_{2,3}$. $A_0$, $A_{11}$, $A_{12}$, and $A_{14}$ are material-dependent parameters describing hopping strengths and mass terms. The interlayer coupling Hamiltonian between neighboring layers is given by:
\begin{eqnarray}
H_z &=& \left(
\begin{array}{cccc}
 0 & t_z &  &  \\
 t_z^\dagger & \ddots & \ddots &  \\
  & \ddots & \ddots & t_z \\
  &  & t_z^\dagger & 0
\end{array}
\right)~,
\nt
\end{eqnarray}

\begin{equation}
t_z = t_0 I_4 + \sum_{i=1}^5 t_i \Gamma_i, \nt
\end{equation}
where:
\begin{gather}
%\begin{aligned}
%\label{eq:t-defs}
t_0 = B_0 \sum_{j=1}^{3} e^{ i \mathbf{k} \cdot \mathbf{b}_j}, \nt \\
t_1 = -i B_{14}\Big( e^{ i \mathbf{k} \cdot \mathbf{b}_1}       + \cos\Omega  [ e^{ i \mathbf{k} \cdot \mathbf{b}_2} + e^{ i \mathbf{k} \cdot \mathbf{b}_3} ] \Big), \nt \\
t_2 = i B_{14}\sin\Omega  \big( e^{ i \mathbf{k} \cdot \mathbf{b}_2}
      - e^{ i \mathbf{k} \cdot \mathbf{b}_3} \big),~~t_3 = 0,\nt \\
t_4 = i B_{12} \sum_{j=1}^{3} e^{ i \mathbf{k} \cdot \mathbf{b}_j}, ~~ t_5 = i B_{11} \sum_{j=1}^{3} e^{ i \mathbf{k} \cdot \mathbf{b}_j}.\nt 
%\end{aligned}
\end{gather}
The lattice vectors connecting layers in an ABC like rhombohedral packing as in Bi$_2$Se$_3$ are (Fig.~\ref{fig:01}):
\begin{equation} \label{eq:reciprocal-vectors}
\begin{split}
\mathbf{b}_1 &= \left( 0, \tfrac{\sqrt{3}a}{2}, c \right), 
\mathbf{b}_2 = \left( -\tfrac{a}{2}, -\tfrac{\sqrt{3}a}{6}, c \right), \\
\mathbf{b}_3 &= \left( \tfrac{a}{2}, -\tfrac{\sqrt{3}a}{6}, 0 \right)
\end{split}
\end{equation}
with $B_0$, $B_{11}$, $B_{12}$, and $B_{14}$ describing interlayer hopping and coupling strengths, and $c$ being the lattice constant along the stacking ($z$) direction.
Figure \ref{fig:01} shows the band structure of the TI slab \cite{ebihara2012finite} along the high-symmetry $-\bar{M}-\Gamma-\bar{M}$ direction, where the Dirac cone-like topological surface states are evident. Vertical interband transitions to conduction bands are associated with changes in spin orientation ($\Delta S_y = \pm 1$), which enable helicity-selective optical excitation and thus contribute to the CPGE. The spin asymmetry of these transitions is the key mechanism enabling net injection current under circularly polarized illumination. Constant-energy contours in Fig.\ref{fig:01} show weak warpings due to the small conduction band mass of Bi$_2$Se$_3$, evident  at $E = 0.6$ eV, while the heavy valence band shows pronounced warping at $E = -0.6$ eV.

\begin{figure}[t!] 
  \centering
\includegraphics[width=0.55\columnwidth]{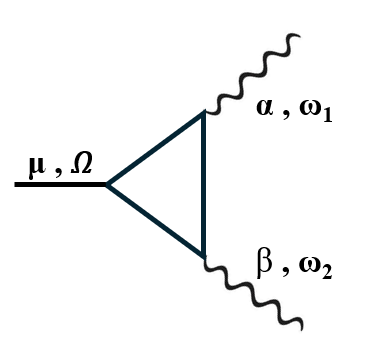} % or .png/.jpg
  \caption{Diagrammatic representation of the injection current contribution to CPGE: the generated photocurrent $j_\mu(\Omega)$ couples to two optical vertices $\alpha,\omega_1$ and $\beta,\omega_2$, giving  $j_\mu(\Omega)= \sigma^\Delta_{\mu\alpha\beta}(\omega_1,\omega_2) E_\alpha(\omega_1)E_\beta(\omega_2)$.}
  \label{fig:2}
\end{figure}

\subsection{Nonlinear Optical Conductivity}
The second order nonlinear photocurrent is a superposition of Fermi surface, shift and injection current contributions \cite{Xie2025}. Together these contributions generate current densities that depend on the stimulating electric field intensity, as well as its optical linear and circular polarizations. Generally, the nonlinear photocurrent $j_\mu$ is expressed in terms of the second-order conductivity tensor connecting to the applied optical field components: 
\begin{equation}
    j_{\mu}(\Omega) = \sum_{\alpha,\beta} \sigma_{\mu \alpha \beta}(\omega_1,\omega_2) 
    { E}_{\alpha}(\omega_1) { E}_{\beta}(\omega_2)~,
\end{equation}
where $\sigma_{\mu \alpha \beta}$ is the nonlinear optical conductivity tensor and ${ E}_{\alpha}(\omega_1)$ and ${ E}_{\beta}(\omega_2)$ are the Fourier components of the electric field. The optical field can be written in real space as \cite{boyd2008nonlinear}:
\begin{eqnarray}
    \mathbf{{E}}(t) = {E}_0   \hat{e}  e^{-i\omega t} +{E}_0^*   \hat{e}^*   e^{+i\omega t}~,
\end{eqnarray}
with amplitude ${E}_0$ and polarization unit vector $\hat{e}$. Typically, however, experiments will measure photocurrent $j$ as a function of quarter wave plate angle $\alpha$ that controls the helicity of light and report photocurrents of the form: 
\begin{equation}
\label{jalpha}
    \frac{j_x}{{ E}_0^2} = C \sin(2\alpha) + L_1 \sin(4\alpha)  +  L_2 \cos(4\alpha)  +  D ~,    
\end{equation}
where the $\sigma_{\mu\alpha\beta}$ elements have been absorbed into the coefficients $C$, $L_{1,2}$ and $D$ which respectively represent circular polarization, linear polarization, and polarization-independent responses, and  $\alpha$ parametrizes the polarization state of light {(see derivation in Appendix A and Eq.~\eqref{eq:A8})}.

In this work, we are primarily interested in the response to circular polarization, which for Dirac states and TIs has been shown to be dominated by the injection current due to spin-momentum locking \cite{Avdoshkin2020}. Physically, the injection current arises from the asymmetric excitation of carriers into the conduction band, where optical selection rules preferentially populate states at specific crystal momenta. This imbalance produces a net current that grows linearly in time while the light is applied. In the Feynman diagrammatic description, the process is represented by two photon-electron interaction vertices on a single electron propagator \cite{Parker2019,Nakazawa2025}. The two field insertions signal the second order character: one interaction promotes an electron from the valence to the conduction band, and the other encodes interference between distinct excitation pathways that generate an asymmetric distribution of conduction--band populations (Fig. \ref{fig:2}). In contrast to the shift current, which comes from coherent interband polarization, the injection current reflects a genuine distribution effect, diagrammatically captured by the asymmetric occupation of final states following photon absorption.

Returning to Eq. \eqref{jalpha}, we would like to identify the tensor elements that correspond to $C$ and $L_{1,2}$ in the case of the circular and linear photogalvanic effect (CPGE and LPGE) respectively. In Appendix A we present the standard symmetry analysis for $\sigma_{\mu\alpha\beta}$ using the C$_{3v}$ point group symmetry  for the surface of Bi$_2$Se$_3$, and the fact that for photocurrent generation $\sigma_{\mu\alpha\beta} = \sigma_{\mu \beta \alpha}^*$, to arrive at the simple relations \cite{connelly2024emergence,hsieh2011nonlinear,Tanaka2023}:
\begin{equation}
j_{x,\mathrm{CPGE}} \propto \Im \sigma_{xxz} ~,~ 
j_{x,\mathrm{LPGE}} \propto \Re \sigma_{xxy}~.\notag
\end{equation}
Following previous works to derive the Feynman diagrammatic method of calculating $\sigma_{\mu\alpha\beta}$, we limit ourselves to the injection current represented by the diagram in Fig. \ref{fig:2}, and the corresponding restricted conductivity $\sigma^\Delta$ \cite{Parker2019,Avdoshkin2020,han2024design}. This diagram is computed as:
\begin{widetext}
\begin{eqnarray}
\label{eq:sigma-abc}
\sigma^{\mu\alpha\beta}_\Delta(\omega_a,\omega_b) 
&=& \frac{- ie^3/\hbar^2}{\pi^{2} \omega_{n_1} \omega_{n_2}}\times \nt\\
&\!\!\!& \!\!\!\!\!\!\!\!\!\!\!\!\!\!\!\!\!\!   \int d\mathbf{k} \sum_{\omega_l}
   \mathrm{Tr}\Bigl[
   v_\mu(\mathbf{k}) G(\mathbf{k},i\omega_l) 
 v_\alpha(\mathbf{k}) G(\mathbf{k},i\omega_l+i\omega_{n1})  v_\beta(\mathbf{k}) G(\mathbf{k},i\omega_l+i \omega_{n1}+i\omega_{n2})
   \Bigr]_{ \displaystyle i\omega_{n1(2)} \to \omega_{a(b)} + i\delta }~.
\end{eqnarray}
which is equivalent to the Kubo formula presented in previous texts \cite{han2024design}. Here $v_\alpha$ is the velocity operator and $G(\mathbf{k},i\omega)$ is the matrix Green's function, q is the charge of the electron, k the absolute value of the momentum, $T$ is the temperature with $\omega_n = 2n\pi T$ and
$\omega_l = (2n+l)\pi T$ the Boson and Fermion Matsubara frequencies, $n$ and $l$ are integers and Tr is a trace. To obtain the
nonlinear conductivity, which is a real frequency quantity, we analytically continue from imaginary
$i\omega_n$ to real $\omega$, with $\delta$ a broadening parameter (for the long wavelength limit $q\to\ 0$). In this work we generally assumed $\delta$=0.005 eV. 
The integral over momentum space is given by 
$\int d {\bf{k}} = \int {d k_x   d k_y}/{4\pi^2}$ for two dimensions (2D).
The matrix Green’s function $ \mathbf{G}(\mathbf{k}, i\omega_l) $ can be conveniently written in terms of the matrix spectral function $ \mathbf{A}(\mathbf{k}, \omega) $ as: 
\begin{equation}
\mathbf{G}( \mathbf{k}, i\omega_l) =
\int_{-\infty}^{\infty} \frac{d\omega}{2\pi} 
\frac{\mathbf{A}(\mathbf{k}, \omega)}{i\omega_l - \omega}~.
\end{equation}
Using this form of the Green function, we can eliminate the frequency sums in $\sigma_{\alpha\beta\gamma}$ and arrive at the expression:
\begin{eqnarray}
\sigma_{\mu\alpha\beta}^\Delta(\omega_1, \omega_2) &=&  \frac{e^3/\hbar^2}{(\omega_1 + i\delta)(\omega_2 + i\delta)} 
\times \notag\\
&~& \int d k \sum_{ijk} \frac{V_{ki}^\mu V_{ij}^\alpha V_{jk}^\beta}{\omega_1 + \omega_2+  i 2 \delta - E_k(\mathbf{k}) + E_i(\mathbf{k})}   \left[
\frac{f_D(E_i(\mathbf{k})) - f_D(E_j(\mathbf{k}))}{\omega_1 + i\delta + E_i(\mathbf{k}) - E_j(\mathbf{k}) }
-
\frac{f_D(E_j(\mathbf{k} )) - f_D(E_k(\mathbf{k}))}{\omega_2 + i \delta + E_j(\mathbf{k}) - E_k(\mathbf{k})}
\right]~.\notag\\
\label{sigmadirty}
\end{eqnarray}
\end{widetext}
The function $f(\omega)$ represents the Fermi-Dirac distribution, which is given by:
\begin{equation}
f_D(\omega) = \frac{1}{e^{(\omega - E_f)/k_B T} + 1}
\end{equation}
where $E_f$ is the Fermi level, $k_B$ is the Boltzmann constant, and $T$ is the temperature. The parameters $\omega_{1,2}$ denote incident frequencies, typically related to external perturbations or applied electromagnetic fields.
$V_{ki}^\alpha$, $V_{ij}^\beta$, and $V_{jk}^\gamma$ are matrix elements computed using eigenvectors $\ket{i}$ with eigenvalues $E_i(\mathbf{k})$:  
\begin{equation}
 V_{ki}^{\alpha} = \langle k | v_{\alpha}(\mathbf{k}) | i \rangle ~.
\end{equation}
We use the tight binding model to compute velocity operators in the planar directions $(\mathbf{k}) =(k_x,k_y)$ as:
\begin{equation}
v_{\alpha}(\mathbf{k}) =  \frac{\partial H(\mathbf{k})}{\hbar \partial k_{\alpha}}~.
\end{equation}
The real space $z$-direction of the tight binding model along the growth direction between slab planes requires the $z$ velocity operator to be computed in real space as:
\begin{equation}
v_{z}(\mathbf{k}) =  - \frac{i}{\hbar} \left[\hat{z},H(\mathbf{k}) \right]~.
\end{equation}
The $z$-position operator $\hat{z}$ for the tight binding model acts on the slab layer (i.e. TI quintuple layer). For $N_z$ layers it is the block diagonal matrix:
\begin{eqnarray}
    \hat{z} = \left(\begin{array}{ccc} 1& & \\ &\ddots & \\ & & N_z \end{array}\right) \otimes \mathbb{1}_{4\times4}~.
\end{eqnarray}
The expression in Eq. \eqref{sigmadirty} can be written in a much more compact form that better lends itself to numerical simulation. First, we note that two sets of triangle diagrams with reversed vertices contribute to the injection current, and thus the \textit{effective} injection current conductivity tensor element is: 
\begin{eqnarray}
    \sigma^\text{inj}_{\mu\alpha\beta}(\omega_1,\omega_2) \equiv \sigma^\Delta_{\mu\alpha\beta}(\omega_1,\omega_2)+\sigma^\Delta_{\mu\beta\alpha}(\omega_2,\omega_1)~.
\end{eqnarray}
We define the tensor of eigen-energy differences and Fermi-Dirac occupation differences respectively as $\varepsilon_{mn} = E_m(\mathbf{k}) - E_n(\mathbf{k})$ and $f_{mn}=f_D(E_m(\mathbf{k})) - f_D(E_n(\mathbf{k}))$. We note the sum frequency $\omega = \omega_1 + \omega_2$, suppresses imaginary offsets (i.e. $\omega_1 \equiv \omega_1 + i \delta $), and then construct the tensors:
\begin{gather}
(t_0^{\mu})_{ij} = \frac{V^{\mu}_{ij}}{\Omega - \varepsilon_{ij}}~,~ (t_1^{\alpha})_{ij} = \frac{f_{ij} V^{\alpha}_{ij}}{\omega_1 + \varepsilon_{ij}}~,\nt\\
(t_2^{\beta})_{ij} = \frac{f_{ij} V^{\beta}_{ij}}{\omega_2 + \varepsilon_{ij}}~.    
\end{gather}
We have now defined $\Omega = \omega_1 + \omega_2$. The effective injection current can then be compactly expressed as: 
\begin{equation}
\label{sigmaclean}
\sigma_{\mu \alpha \beta }^{\text{inj}} =  \frac{e^3/\hbar^2}{\omega_1 \omega_2}   \text{Tr} \left[ t_0^{\mu} \left( t_1^{\alpha} V^{\beta} + t_2^{\beta} V^{\alpha} - V^{\alpha} t_2^{\beta} - V^{\beta} t_1^{\alpha} \right) \right].
\end{equation}
Here the Tr operator sums over every degree of freedom such as tensor indices, eigenvalues and momentum. The compact form in Eq.~\eqref{sigmaclean} is particularly convenient for numerical evaluation of the injection current contribution, which originates from virtual three-band processes encoded by the triangle diagrams. 

{In addition to the injection current, the nonlinear optical response also contains a two-band contribution corresponding to the shift current. The shift current contribution arises from the two-band sector of Eq.~(\ref{eq:A6}) and is given by:}
{\begin{equation}
\begin{split}
\sigma_{\mu \alpha \beta}^{\mathrm{shift}}
(\omega; \omega_1, \omega_2)
&=
\frac{-e^3}{\hbar^2 \omega_1 \omega_2}
\sum_{a,b}
\int [d\mathbf{k}]
\Bigg[
f_{ab}
\frac{
v^\alpha_{ab}
h^{\mu \beta}_{ba}
}{
\omega_1 - \varepsilon_{ab}
}
\\
&\quad+
f_{ab}
\frac{
\tfrac{1}{2}
h^{\alpha \beta}_{ab}
v^\mu_{ba}
}{
\omega - \varepsilon_{ab}
}
\\
&\quad+
[(\alpha,\omega_1)\leftrightarrow(\beta,\omega_2)]
\Bigg].
\end{split}
\label{eq:shift_current}
\end{equation}}

{Here, $\varepsilon_{ab}=E_a(\mathbf{k})-E_b(\mathbf{k})$, while $h^{\mu\beta}$, and $h^{\alpha\beta}$ denote covariant derivatives of the velocity operator (explicitly defined in Appendix B). Physically, the shift current describes the real-space displacement of an electronic wave packet during an interband optical transition and is fundamentally connected to the Berry-phase geometry of Bloch states. Unlike the injection current, which originates from asymmetric carrier injection processes, the shift current is governed primarily by coherent two-band interband transitions. Additional details regarding the derivation and numerical implementation of Eq.~\eqref{eq:shift_current} are provided in Appendix~B.}
\begin{figure}[!t]
  \centering
  \includegraphics[width=1\columnwidth]{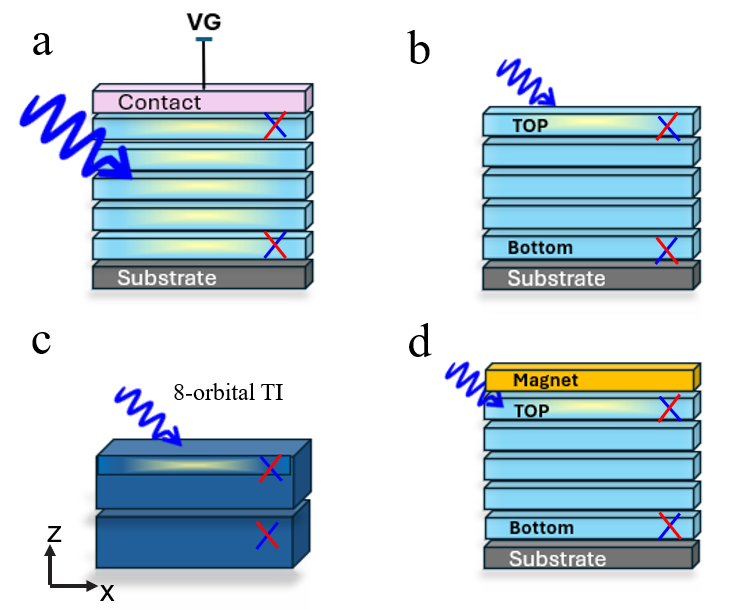}%
  \caption{Schematic of different configurations used to analyze the circular photogalvanic effect (CPGE), (Im $\sigma_{xxz}$) in TIs. (a) Fully irradiated TI slab, (b) TI surface irradiation, (c) 8-orbital TI, (d) TI slab with a magnetic layer applied to the top surface to break time-reversal symmetry.}
  \label{fig:3}
\end{figure}

\subsection{Layer Resolved Injection Current}

The inversion symmetry of Bi$_2$Se$_3$ guarantees that the second order response tensor is actually zero as $\mathbf{I}:\sigma_{\mu \alpha \beta} \mapsto - \sigma_{\mu \alpha \beta}$ under inversion $\mathbf{I}$. However, light does not uniformly penetrate a thin film, and a field that decays into the material as $E = E_0 e^{-\lambda z}$ with absorption coefficient $\lambda$ contributes exponentially diminished weight to the nonlinear photocurrent at each layer. The vertical optical absorption coefficient in Bi$_2$Se$_3$ ranges from $\sim 1/5$ nm$^{-1}$ to $\sim 1/100$ nm$^{-1}$ for visible to long wave infrared light respectively \cite{Yu2023}. This implies that for the typical thin film thicknesses of $\mathcal{O}(10\text{ nm})$ in nonlinear photocurrent generation studies in Bi$_2$Se$_3$, we can expect the bottom surface to be weighted as little as $13 \%$ relative to the top surface. Thin film/substrate interface effects can further hamper the bottom surface's participation in $\sigma_{\mu \alpha \beta}$, so we are motivated to study the layer resolved contributions to the photocurrent \cite{pan2017helicity,Junck2013,huang2021optical}. In particular we would like to extract the impact of the top two ($l=1,2$) and bottom two layers ($l=N_z-1,N_z$) to capture the majority of the topological surface state wavefunctions' contribution to $\sigma^\text{inj}_{\mu\alpha\beta}$.

We can account for the layer resolved contributions to the injection current by separating layer and spin/orbit degrees of freedom as $l=1,\ldots,N_z$ and $s=1,2,3,4$. We observe that the velocity tensor $[v_z(\mathbf{k})]_{({l_1},s_1),({l_2},s_2)}$ encodes the hopping or `current' that flows from layer $l_2$ to $l_1$ \cite{Cao2022,Mahan}. As we care about the current that flows to a given layer $l$, we can compute this via left projection:
\begin{eqnarray}
    v_{z,l}(\mathbf{k}) &\equiv& P_l \cdot v_z(\mathbf{k}) \nt \\
    &=& \text{diag}(0,\dots,0,\underarrow[l \text{th position}]{1},0,\ldots,0)\otimes\mathbb{1}_{4\times4} \cdot v_z(\mathbf{k})~, \nt\\
\end{eqnarray}
where only the $l$th entry of the diagonal matrix is non-zero. By inserting these projectors $P_l$ into our original triangle diagram we can obtain the layer restricted contribution to the injection current:
\begin{widetext}
\begin{eqnarray}
\sigma^\Delta_{\mu\alpha\beta,l}(\omega_a,\omega_b)
&=& \frac{- i e^3/\hbar^2}{\pi^{2} \omega_{n_1} \omega_{n_2}}
   \times\nt\\
   &~& \!\!\!\!\!\!\!\!\!\!\!\!\!\!\!\!\!\!\! \int d\mathbf{k} \sum_{\omega_l}
   \mathrm{Tr}\left[
   v_\mu(\mathbf{k}) G(\mathbf{k},i\omega_l) 
   (P_l v_\alpha(\mathbf{k}) )G(\mathbf{k},i\omega_l+i\omega_{n1})  (P_l v_\beta(\mathbf{k})) G(\mathbf{k},i\omega_l+i \omega_{n1}+i\omega_{n2})
   \right]_{ i\omega_{n1(2)} \to \omega_{a(b)} + i\delta }~.\nt\\
\end{eqnarray}
\end{widetext}
We note that due to the block tri-diagonal form of $G(\mathbf{k},i\omega)$ and $v_\alpha(\mathbf{k})$ the product of the last two terms in the trace above would be zero if non-identical projection layers were used, and therefore  $\sum_{l} \sigma^\Delta_{\mu\alpha\beta,l}(\omega_a,\omega_b) = \sigma^\Delta_{\mu\alpha\beta}(\omega_a,\omega_b)$. 

As a final step we can conduct analysis identical to the previous subsection to obtain a form like Eq. \eqref{sigmaclean} using layer projected velocity tensors. For later convenience as a last step we define the top and bottom injection conductivity terms in the DC limit ($\omega_1 = \omega + i \delta$, $\omega_2 = -\omega + i \delta$) as:
\begin{eqnarray}
    \sigma^T_{\mu\alpha\beta}(\omega) &\equiv& \Im \sigma^\text{inj}_{\mu\alpha\beta,1}+\Im \sigma^\text{inj}_{\mu\alpha\beta,2}~, \nt\\
    \sigma^B_{\mu\alpha\beta}(\omega) &\equiv& \Im \sigma^\text{inj}_{\mu\alpha\beta,N_z-1}+\Im \sigma^\text{inj}_{\mu\alpha\beta,N_z}~. \nt\\   
\end{eqnarray}

\section{Bulk Calculations and Field Dependence}
In this section, we analyze the voltage-tunable nonlinear optical response and polarization-resolved photocurrent generation in our TI slab geometry. Previous experimental studies have observed a strong dependence of CPGE to applied voltages in silicon nanowires \cite{dhara2015voltage} and TIs \cite{huang2021optical}. To illustrate the physical mechanism underlying this voltage-tunable photocurrent response, we first consider the case of uniformly optically excited TIs (as might be the case for extremely thin films, or long wave length excitations), where an applied voltage explicitly breaks the inversion symmetry to mimic contact or gating field effects.

When a finite voltage, $V_G$, is applied to the system as shown in Fig. \ref{fig:3}(a), the slab Hamiltonian explicitly breaks inversion symmetry through the layer dependent potential:
\begin{eqnarray}
    H_V = \text{diag}\left(-\frac{V_G}{2},-\frac{(N_z-1)V_G}{2 N_z},\ldots,\frac{V_G}{2}\right) \otimes \mathbb{1}_{4\times4}~.\nt\\
\end{eqnarray}
In this case, the degeneracy of the surface states is lifted and the surface electronic structure is altered as a gap forms (see Fig. \ref{fig:B_2}). Such behavior is a characteristic of a breaking of the inversion symmetry, as to be expected and especially relevant in fields of optical response under second order, as is the case with the circular photogalvanic effect \cite{dhara2015voltage}.

To demonstrate the impact on the net nonlinear photocurrent generation, we evaluate 
the injection conductivity components $\sigma^{\text{inj}}_{xxz}$ and 
$\sigma^{\text{inj}}_{xxy}$ at a Fermi level of $E_f = 0.2~\mathrm{eV}$ above the Dirac point. This value reflects the typical $n$-type doping of Bi$_2$Se$_3$, where the Fermi level lies within the conduction band due to selenium vacancies, as observed in angle-resolved photoemission spectroscopy (ARPES) measurements~\cite{xia2009observation}. 

\begin{figure*}[!t] 
  \centering
  
  \includegraphics[width=1\linewidth]{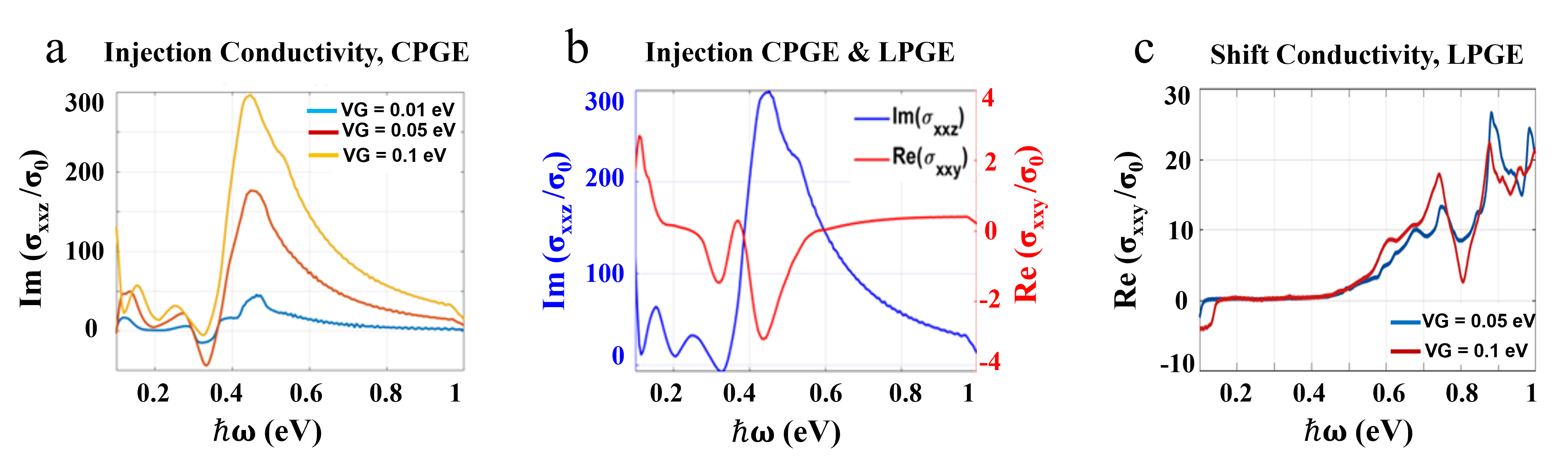} % or .png/.jpg

 \caption{a) imaginary part of the second-order injection conductivity
$\mathrm{Im}(\sigma_{xxz})$ of the bulk material as a function of photon
energy $\hbar\omega$ for different applied surface potentials
$V_0 = 0.01$, $0.05$, and $0.10$ eV. Increasing the applied field strength
significantly enhances the nonlinear optical response, particularly in the
$0.4$--$0.6$ eV spectral range.
b) comparison between the CPGE injection conductivity
$\mathrm{Im}(\sigma_{xxz})$ (blue axis) and the LPGE injection conductivity
$\mathrm{Re}(\sigma_{xxy})$ (red axis) for $V_0 = 0.10$ eV, showing that
the CPGE response dominates at low photon energies while becoming comparable
to the LPGE response at higher energies. 
c) Real part of the LPGE shift conductivity
$\mathrm{Re}(\sigma_{xxy})$ as a function of photon energy for
$V_0 = 0.05$ and $0.10$ eV. The LPGE response increases strongly above
$\sim 0.6$ eV and exhibits pronounced spectral features near
$0.8$--$1.0$ eV.
Note: the sheet conductivity conversion factor is $\sigma_0 \approx 5.90 \times 10^{-15}~\mathrm{A\,m/V^2}$. }
  \label{fig:4}
\end{figure*}

%%%%%%%%%%%%%%
Figure \ref{fig:4}(a) shows how $\Im \sigma^\text{inj}_{xxz}$ increases from 0 as $V_G$ is applied, consistent with $\sigma_{\mu\alpha\beta} \neq 0$ when inversion symmetry is broken. {At low applied gate voltages ($V_G \sim 0.01$), the nonlinear response remains comparatively weak. The reason is weak inversion symmetry breaking and the limited asymmetry in the band structure near the Dirac point, which restricts the phase space and matrix elements contributing to the injection current. As $V_G$ increases, the enhanced symmetry breaking and band distortion lead to a stronger nonlinear optical response.} The observed peaks in $\Im \sigma_{xxz}$ are attributed to resonant interband transitions that coincide with regions of high density of states and pronounced Berry curvature effects, particularly in the photon energy range of 0.3 to 0.6 eV. Our results clearly demonstrate that the second-order nonlinear optical conductivity component $\mathrm{Im}(\sigma_{xxz})$ exhibits significant resonances with photon energy, whose amplitude and spectral position are highly sensitive to the applied external gate voltage. These resonances originate from field-induced modifications of the band structure, including Dirac cone deformation, band hybridization, and increased asymmetry. 
Figure \ref{fig:4}(b) compares the CPGE injection conductivity, $\mathrm{Im}\,\sigma_{xxz}$, and the LPGE injection conductivity, $\mathrm{Re}\,\sigma_{xxy}$, at $V_G = 0.1$ eV. { For this model of Bi$_2$Se$_3$ we find across photon energy spectrum that the CPGE response strongly dominates ($\mathrm{CPGE} \gg \mathrm{LPGE}$).}

{ In Appendix B we discuss how weak warping near the Dirac point of Bi$_2$Se$_3$ reduces the injection contribution to the LPGE \cite{Li2019}. For DC photocurrent generation ($\omega_1 = \omega + i \delta, \omega_2 = - \omega + i\delta$), the resonant $1/(2i \delta - \varepsilon_{ij})$ in Eq. \eqref{sigmadirty} and weak warping ultimately results in a leading order CPGE to LPGE injection current ratio of:
\begin{eqnarray}
    \frac{\text{CPGE}^{\text{inj}}}{\text{LPGE}^{\text{inj}}} \sim \frac{E_g}{\delta} \sim 100~,
\end{eqnarray}
where for our simple tight binding model, which lacks many-body corrections, the band gap is $E_g \sim$ 0.4 eV \cite{Aguilera2019}. This simple approximation affirms the relative weight of CPGE and LPGE in Fig. \ref{fig:4}. For other topological insulators where warping is more pronounced, like Bi$_2$Te$_3$, we would expect a deviation from this approximation \cite{fu2009hexagonal}.

The dominant contribution to LPGE is thus expected to originate from the shift current, which we detail how to calculate in Appendix B.  Figure \ref{fig:4}(c) shows the real part of the LPGE shift conductivity, $\mathrm{Re}\,\sigma_{xxy}$, as a function of photon energy for $V_G = 0.05$ and $0.10$ eV. In contrast to the strongly resonant CPGE injection response, the LPGE shift conductivity evolves more gradually with photon energy and becomes significant mainly above the band gap, peaking at $\sim 0.6$ eV. 

Overall we can conclude that at low photon energies the CPGE response strongly dominates ($\mathrm{CPGE} \gg \mathrm{LPGE}$), while at higher photon energies the two responses become comparable in magnitude, consistent with experimental observations in topological insulators \cite{huang2021optical}. The pronounced spectral structures near $0.8$--$1.0$ eV originate from higher-energy interband transitions enhanced by gate-induced symmetry breaking and band-structure anisotropy.
}

Our calculations thus far suggest that an applied field controls not only the strength of the photocurrent (shown in Fig. \ref{fig:B_3}), but also the relative polarization dependencies. Phenomenologically the applied field term $H_V$ can arise due to gating effects or contact field effects, which in both cases have shown to lead to enhanced CPGE to LPGE ratios \cite{pan2017helicity,dhara2015voltage}. Our findings suggest a clear strategy for engineering high-sensitivity photodetection, utilizing resonant frequencies and strong-field-induced nonlinear effects to optimize device performance. The analysis so far has assumed uniform absorption throughout the TI. The next section presents the corresponding layer-resolved calculation, which provides a more complete description.

\section{Layer-Resolved Computation}
\subsection{Four Band Model}
In this section, we perform layer-resolved calculations of the injection components of the circular photogalvanic effect (CPGE) in TIs using the slab model in Section II with tight binding parameters corresponding to Bi$_2$Se$_3$ as presented in Appendix C and 15 layers. We focus on the imaginary part of the nonlinear injection conductivity, $\operatorname{Im}\,\sigma^\text{inj}_{xxz}(\omega)$, which characterizes the helicity-dependent photocurrent along the $x$-direction under circularly polarized light.
A key property of the slab is that the top and bottom surfaces are related by the
mirror symmetry $M_z:(x,y,z)\mapsto(x,y,-z)$, which reverses the surface normal. Because $\sigma_{xxz}$ carries a single $z$ index, it is odd under inversion:
\begin{equation}
  \sigma^{\mathrm{Top}}_{xxz}(\omega) = -\sigma^{\mathrm{Bottom}}_{xxz}(\omega) ~.
\end{equation}
Thus, the top and bottom surface CPGEs have opposite signs due to their opposite spin--momentum locking and the reversal of velocity matrix elements when the surface normal is inverted. For an inversion-symmetric slab, these contributions cancel. Accordingly, in what follows we report results for a single surface (the top surface); the bottom surface is obtained by an overall sign flip.  As mentioned in Section II, in order to capture the majority of the topological surface state wavefunctions, we consider the top two and bottom two layers of the material stack as the `surfaces' in our calculations. In Fig. \ref{fig:D1} (Appendix B) we present $\sigma^{\mathrm{T,B}}_{xxz}(\omega)$ where the top (solid) and bottom (dashed) results for different optical frequencies and Fermi levels are seen to be opposite each other.

Note that in the very low–frequency limit ($\hbar\omega \to 0$), second–order dc photocurrents are numerically sensitive. Within the length–gauge formalism, the injection (helicity–dependent) contribution involves resonant denominators together with a phenomenological relaxation factor scaling as $1/\delta$. 
When combined with the Dirac-like dispersion, the resulting integrands become sharply peaked near the Brillouin-zone center and require extremely dense $k$-meshes and small broadening $\delta$ for convergence to zero at $\hbar \omega =0$. We demonstrate this convergence issue in the Appendix B Fig. \ref{fig:D1}. 
%As illustrated in Fig.~\ref{fig:D1}, the apparent mismatch between top and bottom surface responses at $\hbar\omega \lesssim 0.1~\mathrm{eV}$ mainly reflects numerical convergence noise and residual mixing with intraband (Drude-like) terms, rather than intrinsic physics. 
To avoid over interpretation of this regime and facilitate large parameter sweeps, all quantitative spectra reported in rest of the main text correspond to $\hbar\omega \ge 0.1~\mathrm{eV}$, where the interband-dominated response is well converged with respect to $k$-mesh, slab thickness, and the relaxation parameter $\delta$.

\begin{figure}[h!]
  \centering
  \includegraphics[width=0.8\columnwidth]{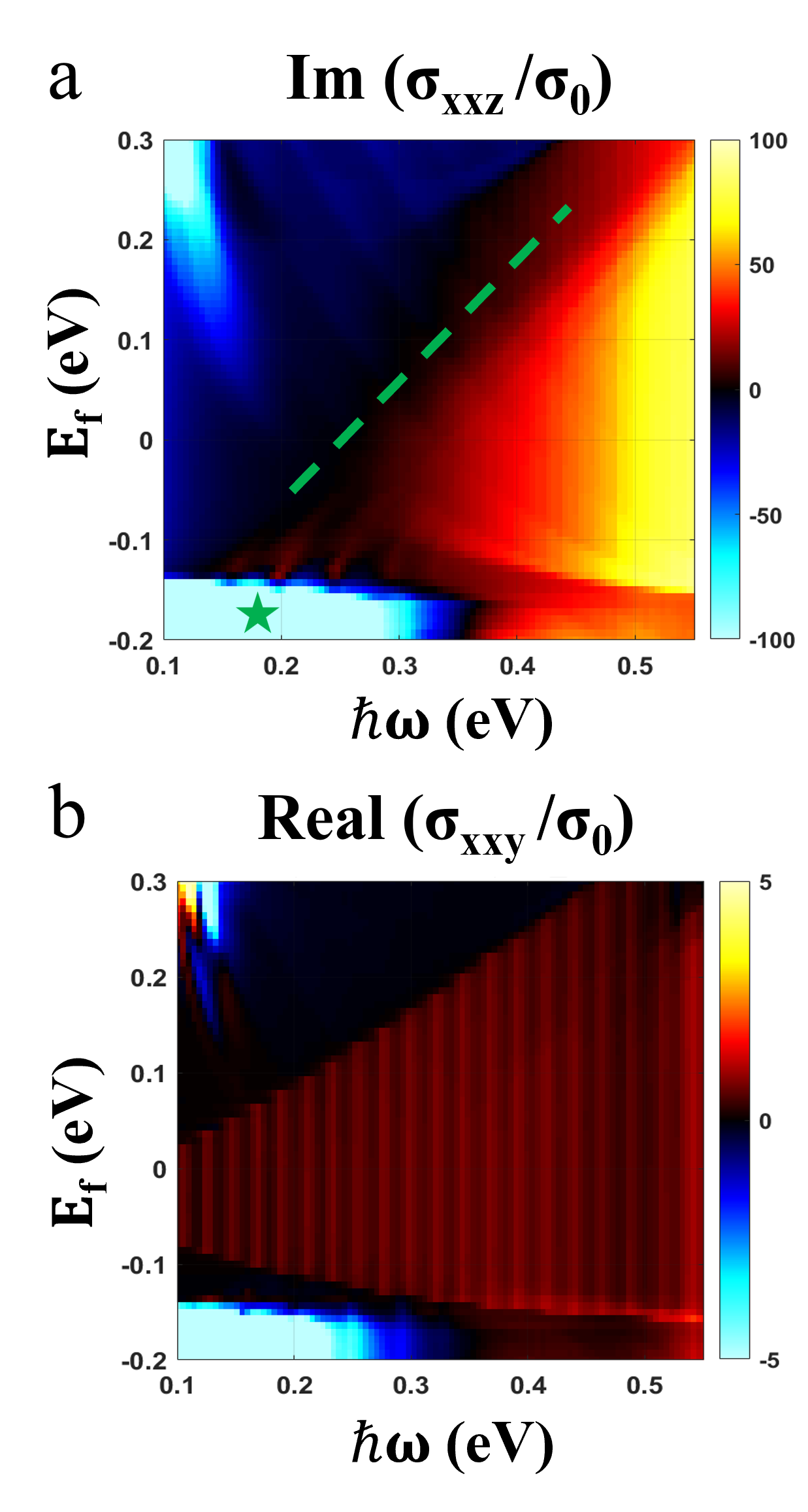}%
  \vfill
  
  \caption{a) Dependence of top surface layer CPGE $\propto \text{Im} (\sigma_{xxz})$ on Fermi level {$E_f$}, and excitation energy, $\hbar \omega$. The surface CPGE flips sign across the green line $E_f = \hbar\omega$. A green star shows that for p-type Bi$_2$Se$_3$, the flattish top of the valence band leads to a large density of states and a corresponding large negative CPGE. b) Plot of LPGE response ($\mathrm{Re} (\sigma_{xxy})$) versus photon energy and Fermi level. LPGE is strongly suppressed relative to CPGE across the parameter space, consistent with surface-cancellation in the slab and the weak in-plane anisotropy of the model. Note: $\sigma_0 \approx 5.90 \times 10^{-15}~\mathrm{A\,m/V^2}$.}
  \label{fig:5}
\end{figure}

 Figure \ref{fig:5}(a) presents a colormap of the calculated $\operatorname{Im} \sigma_{xxz}$ on the top surface as a function of the Fermi level $E_f$ and the photon energy $\hbar\omega$. A pronounced change in the CPGE response is observed across the line $\hbar\omega \sim E_f$, dashed green line, which marks the condition where the photon energy becomes resonant with interband absorption across the Fermi level. The sign reversal of the CPGE across this line indicates a transition from valence-to-conduction to conduction-to-conduction excitation processes, which is consistent with the literature \cite{pan2017helicity}. Further note that there is a finite width to the  region where $\sigma^{\mathrm{T}}_{xxz}(\omega\approx E_f) $, which reflects the well known result that intra-Dirac cone excitations yield zero CPGE unless photon drag is included \cite{Junck2013,Xie2025,plank2016photon}. Once the photon energy is large  enough to connect surface and bulk states, then $\sigma^{\mathrm{T}}_{xxz}(\omega\approx E_f)$ increases from zero. This region of zero conductivity naturally disappears as $E_f$ enters the bulk bands.  In the case of $p$-type Bi$_2$Se$_3$, marked by the green star in Fig. \ref{fig:5}(a), the asymmetry in optical excitation is enhanced due to the relatively flat top of the valence band, resulting in a strong negative CPGE response.

Figure~\ref{fig:5}(b) presents the LPGE response, $\mathrm{Re}\,\sigma^{\text{inj}}_{xxy}$, as a function of Fermi level $E_f$ and photon energy $\hbar\omega$. Across the entire $(E_f,\hbar\omega)$ range, the LPGE magnitude remains much smaller than the corresponding CPGE response, confirming that it does not influence our main conclusions. The injection-type LPGE channel follows the intrinsic $C_{3v}$ symmetry of the topological surface states and arises mainly from hexagonal warping effects—particularly those affecting the valence band—whose strength is weak within this model. The fringe-like structures observed in $\mathrm{Re}\,\sigma^{\text{inj}}_{xxy}$ caused by discrete interband transitions between quantized subbands in the finite slab geometry. These features reflect finite-size quantization of surface states, where each stripe corresponds to a resonant transition condition $E_{n,k}-E_{m,k'}=\hbar\omega$, and their modulation by hexagonal warping underscores the symmetry-allowed yet weak nature of the LPGE response.

\begin{figure}[t]
  \centering
  \includegraphics[width=1\columnwidth]{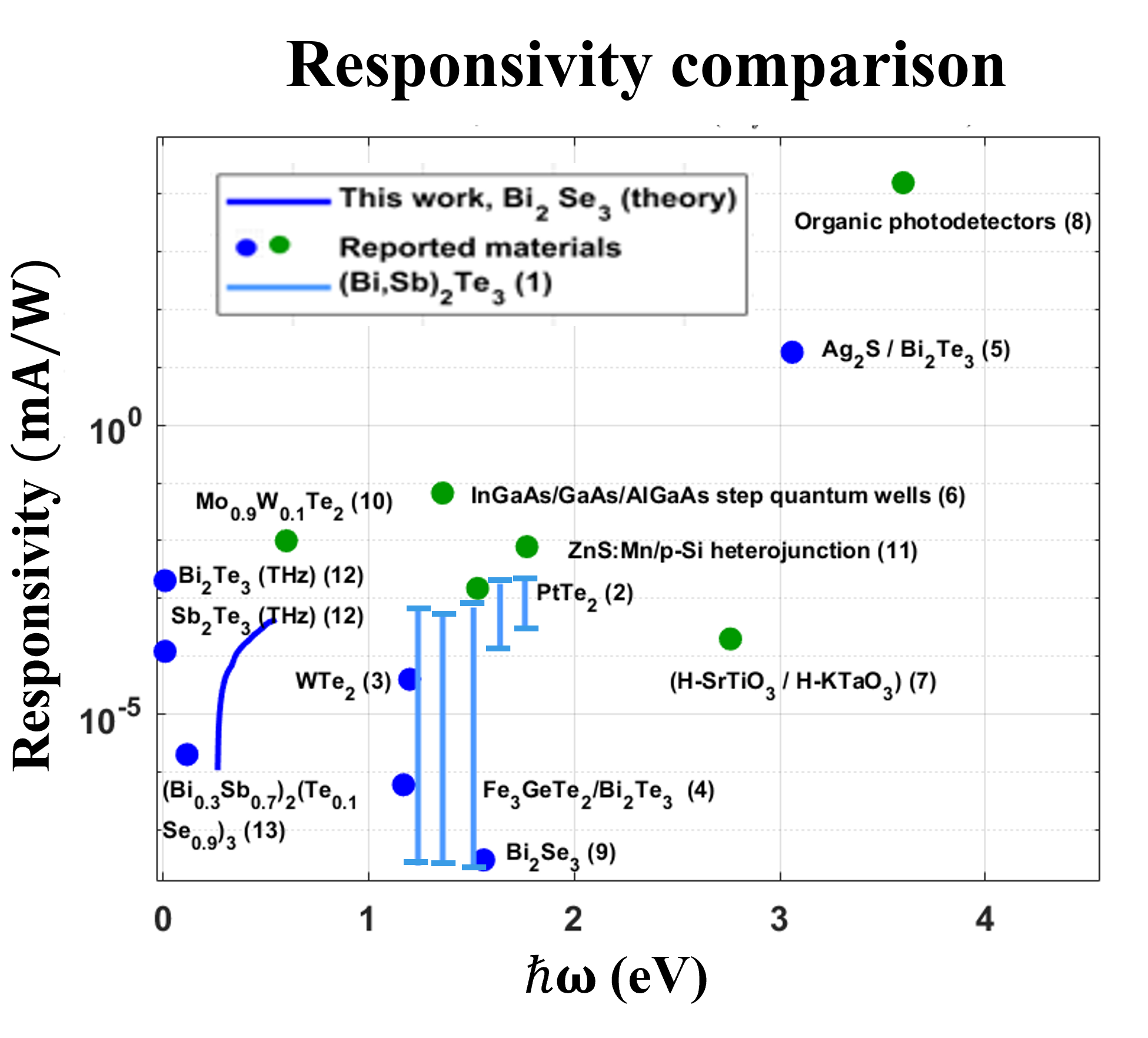}
  \caption{Responsivity versus photon energy for a range of photodetector materials, including conventional semiconductors, topological materials, and emerging 2D systems. The responsivity calculated from Fig. \ref{fig:5}(a) at $E_f=0$ eV is shown by a dark blue line has a competitive regime relative to experimentally reported values especially at longer wavelengths (blue dots and lines represent TIs) . Numbers 1-13 represent experimentally reported responsivities from Refs.~\cite{pan2017helicity,liu2025ptte2,xu2018electrically,you2025interface,song2024high,yu2014spin,li2025circular, liu2022chiral,mciver2012control, ji2019spatially,kumar2023fabrication,olbrich2014room,danilov2021superlinear}, respectively.
}
  \label{fig:6}
\end{figure}

To evaluate the device performance, we estimate the photoresponsivity arising from the CPGE-induced photocurrent. The calculated responsivity originates from the first two layers of the topological insulator and corresponds to illumination by a mid-wave infrared laser with a typical spot diameter of 250~$\mu$m. Using the calculated nonlinear conductivity, we obtain a responsivity of $R = 0.170~\mathrm{\mu A/W}$ (Appendix D).
As Figure \ref{fig:6} shows, this value is comparable to those reported experimentally for high-performance 
topological photodetectors~\cite{olbrich2014room,plank2018review, song2024high}, where other technologies require cooling. {Notably, the chiral surface states provide an additional advantage by enabling multi-spectral functionality: at a fixed Fermi energy, the CPGE response generates opposite photocurrent directions for different spectral bands, allowing a single detector to distinguish between them. It indicates that the circular photogalvanic effect in our system can generate a measurable direct photocurrent under moderate optical excitation, demonstrating efficient light–matter coupling and potential applicability in room-temperature  multi-spectral optoelectronic polarimetric detection. The Fermi level sensitivity of the response provides additional functionality to TI-based detectors, as they can thus be gate tuned to refine their response to a given wavelength of light.}

\subsection{Eight Band Model}

\begin{figure*}[t!]
\centering
\includegraphics[width=2\columnwidth]{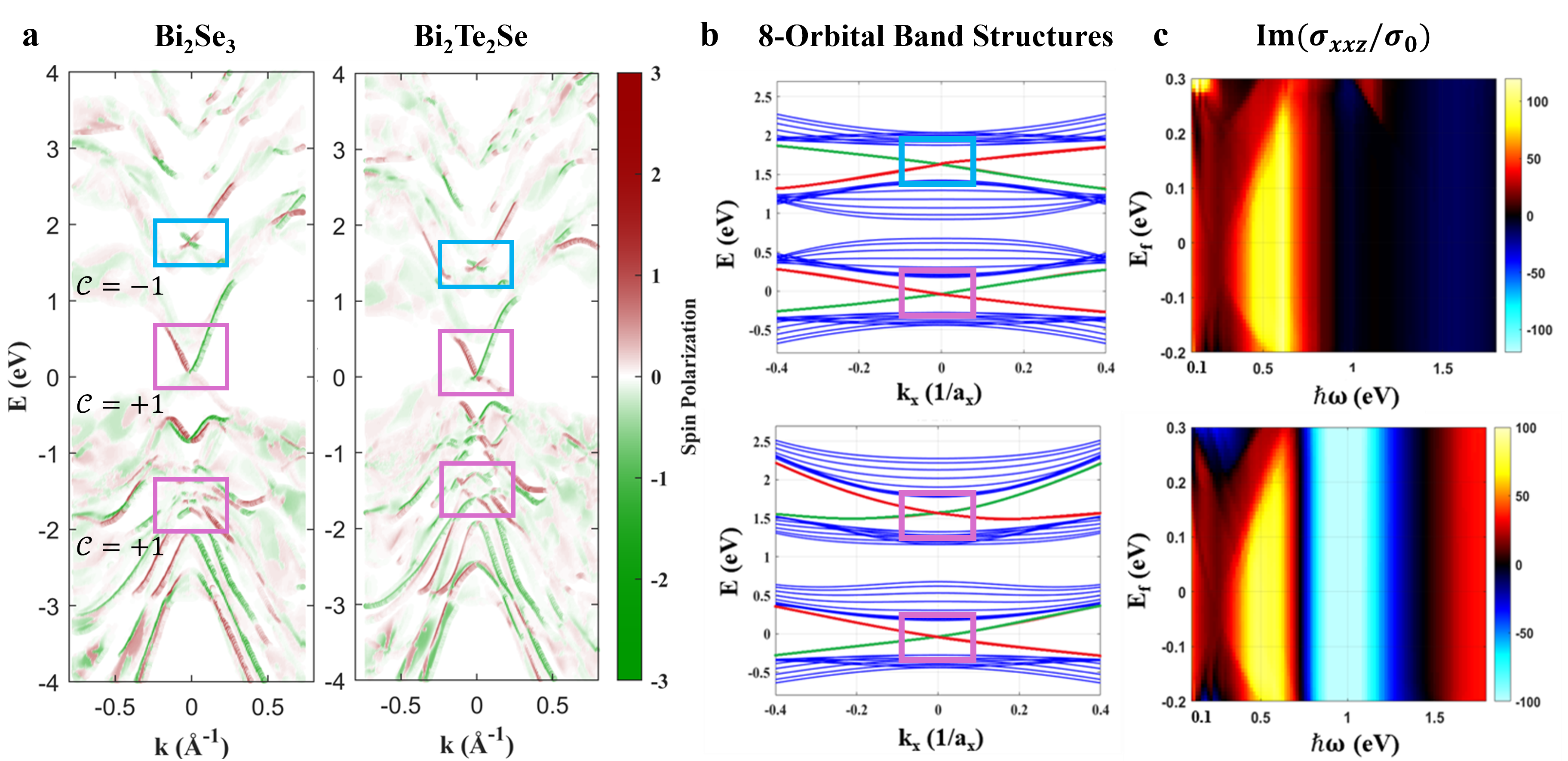}
    \caption{a) Surface projected band structures of Bi$_2$Se$_3$ and Bi$_2$Te$_2$Se using data from Ref. \cite{Aguilera2019}. Multiple surface states exist in the conduction and valence bands with the same and opposite chiralities as the Dirac state in the principal gap. Pink/blue boxes highlight states of chirality $\mathcal{C}=\pm 1$ respectively. b) Simple 8-orbital band structures developed to probe the nonlinear conductivity dependence on transitions between opposite and identical chirality topological bands. c) Calculations for the top surface contribution to CPGE,  $\Im \sigma^T_{xxz}$, corresponding to the band structures in (b). For +/- chirality pairing (top), CPGE weakens for transitions between Dirac states, and nearly vanishes due to cancellations of interband transitions between bands of opposite chirality. %for transitions from the principal Dirac state to the lower cone of the higher energy Dirac band. 
    For +/+ however, CPGE magnitude remains constant across the Dirac point, with the opposite sign for $\Im \sigma_{xxz}$ as in the +/- case. These results demonstrate that CPGE will be enhanced by optical excitation between surface states of the same chirality, and motivates ultraviolet CPGE sutdies of Bi$_2$Te$_2$Se to excite the surface states at $\sim -1.5$ eV below the intrinsic Fermi level to the Dirac states at $1.6$ eV above the Fermi level in the conduction band. Note: $\sigma_0 \approx 5.90 \times 10^{-15}~\mathrm{A\,m/V^2}$.}
    \label{fig:7}
\end{figure*}

Several nonlinear optical experiments on (Bi,Sb)$_2$(Te,Se)$_3$ use visible light ($\hbar \omega \sim 1.6 eV$) to generate photocurrents \cite{connelly2024emergence,pan2017helicity,Braun2016,Tu2017}. At this energy range it is well established that the photocurrents involve interband transitions between bulk states, the principal gap topological surface state and one that exists 1.6 eV higher within the conduction bands \cite{Sobota2012,Aguilera2019}. In order to numerically probe the contribution of these higher energy surface state transitions, we would need to extend the tight binding model of section II to several more orbitals, which is beyond the scope of this paper. Instead, we take a phenomenological approach of creating a `doubled' 8-band Hamiltonian consisting of two copies of the Bi$_2$Se$_3$ $H_{\text{tot}}$ Hamiltonian, shifted in energy and coupled through a symmetry allowed term $d$ to prevent trivial crossings: 
\begin{eqnarray}
    H_{8\times8}^{\mathcal{C}_1 \mathcal{C}_2} &=& \left(
    \begin{array}{cc}
    \mathcal{C}_1 H_{\text{tot}} & d \\
    d^\dagger & 1.6 \text{~eV}+ \mathcal{C}_2 H_{\text{tot}}
    \end{array}
    \right)~,\\
    d &=& \Gamma_2 \left\{ \sin( \mathbf{k}\cdot\mathbf{a}_1 )+\right. \nt\\
    &~& \left.~\cos(\Omega)\left[ \sin( \mathbf{k}\cdot\mathbf{a}_2 )+\sin( \mathbf{k}\cdot\mathbf{a}_3 )\right]\right\}~.
    \label{cceqn}
\end{eqnarray}
We've allowed for the two surface states to have opposite ($\mathcal{C}_1\times\mathcal{C}_2 = -1$) or the same ($\mathcal{C}_1\times\mathcal{C}_2 = 1$) chiralities. In Fig. \ref{fig:7}(a): surface band structures of Bi$_2$Se$_3$ and Bi$_2$Te$_2$Se demonstrate that multiple Dirac surface states exist within the topological insulators extended conduction and valence bands, with both $+1$ and $-1$ chiralities \cite{Aguilera2019}. In Fig. \ref{fig:7}(b) we model band structures using Eq. \eqref{cceqn} for both $+/+$ and $+/-$ pairings of chiralities, using pink and blue boxes to highlight surface states with $\mathcal{C} = \pm 1$. The principal and higher conduction topological surface states of Bi$_2$Se$_3$ and Bi$_2$Te$_3$ have $+/-$ chirality pairing. Identical chirality pairs like $+/+$ are also allowed, for example the valence band and principal topological surface states in Bi$_2$Te$_3$ follow a $+/+$ pairing. The resulting calculations for $\sigma^T_{xxz}$ are shown in Fig. \ref{fig:7}(c). The chirality pairing is seen to make a significant difference in the qualitative behavior. 

For the ${+}/{-}$ case, Fig. \ref{fig:7}(c,top) the top and bottom surfaces exhibit opposite spin-momentum locking, which causes their CPGE contributions to partially cancel when integrated over the full slab. For photon energies just below the Dirac point separation, Pauli-blocking further represses the optical transitions, and the CPGE magnitude decreases—consistent with wavelength dependent measurements \cite{pan2017helicity}. At lower photon energies (\(\hbar\omega \lesssim 0.6~\mathrm{eV}\)), the response increases again, recovering the 4-orbital model behavior when higher energy states are no longer accessible. 

%arising from low-energy channels (e.g., top–bottom hybridization or surface–bulk transitions) that remain helicity-sensitive. 
% and . As the Fermi level on the upper surface is tuned below its Dirac point, interband transitions near that cone become Pauli-blocked ($\hbar\omega \gtrsim 2|E_f - E_D|$), 

For the ${+}/{+}$ chirality configuration, Fig. (\ref{fig:7})(c,bottom), both surfaces exhibit the same spin--momentum locking orientation, so their CPGE contributions add constructively, leading to an overall stronger net response. The resulting $\sigma^T_{xxz}$ is more uniform and robust across a broad range of Fermi levels. Notably, the CPGE in this configuration is less sensitive to the precise position of the Fermi level relative to the Dirac point of the surface states. This behavior indicates that surfaces with matching chirality support more stable helicity-driven photocurrents, consistent with the expectation that the symmetry of spin textures governs the net CPGE response in TI slabs.

\begin{figure}[t] % placement: t=top, b=bottom, h=here, H=require exact (needs float)
  \centering
  \includegraphics[width=0.8\columnwidth]{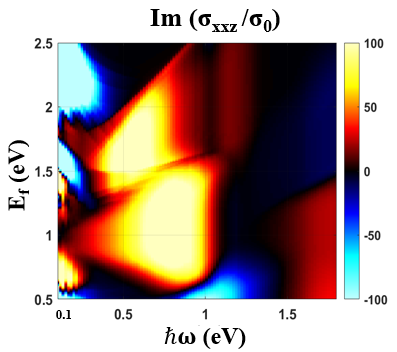} % or .png/.jpg
    \vfill
 \caption{Imaginary part of the second-order nonlinear optical conductivity component $(\sigma^T_{xxz})$ as a function of photon energy $\hbar\omega$ and Fermi level $E_f$, computed for the same $+/+$ configuration. The color map indicates a pronounced nonlinear response in regions associated with second surface state transitions and broken symmetry conditions. Note: $\sigma_0 \approx 5.90 \times 10^{-15}~\mathrm{A\,m/V^2}$.}
  \label{fig:8}
\end{figure}

Figure~\ref{fig:8} presents the $+/+$ configuration colormaps of \(\sigma^T_{xxz}\) as a function of photon energy \(\hbar\omega\) for a different range, \(0.5 < E_f < 2.5~\mathrm{eV}\). By considering larger Fermi levels, we can probe the dependence on transitions between valence and principal topological bands. For photon energies greater than the Dirac point separation ($\hbar \omega > 1.6$ eV), the sign of the helicity dependence flips as the chemical potential is tuned above and below the upper Dirac point at 1.6 eV. Pauli blocking is again manifest in diminished $\sigma_{xxz}$ as photon energies dip below 1.6 eV. However, the additive nature of the $+/+$ chiralities returns for midwave infrared frequencies where robust CPGE of fixed sign persists across a large Fermi level range. 

The comparison of two chirality options  highlights the crucial role of surface-state chirality in determining the strength and tunability of CPGE in topological systems. Understanding and controlling the symmetry and spin texture of surface states offers a promising avenue for engineering tailored photocurrent responses in future optoelectronic and spintronic devices.

\subsection{Magnetic Proximity Effect}
The versatility of a tight binding model allows us to simulate possible engineering and device fabrication approaches to modify the nonlinear photocurrents. To investigate the influence of magnetic interactions on CPGE, we model a TI slab with a magnetic layer deposited on its surface (Fig.~\ref{fig:3}(d)). A magnetic exchange field \(M_z\) is introduced along the \(z\)-axis in the Hamiltonian, breaking time-reversal symmetry and opening a gap at the Dirac point of the surface-state spectrum \cite{vakili2022low}. This magnetic gap shifts the Berry curvature from inside the gap to the region above the edges, altering the circular dichroism of the helicity-selective interband transitions.

Figure \ref{fig:F_1} compares the slab band structures without and with a finite out-of-plane exchange fields. When \(M_z = 0.1\), a clear gap appears at the surface Dirac point, giving rise to a massive Dirac dispersion while leaving the bulk states nearly unchanged. This behavior agrees with the low-energy surface Hamiltonian
\begin{equation}
H_{\mathrm{surf}} = v_F (k_x\sigma_y - k_y\sigma_x) + M_z \sigma_z ,
\end{equation}
where the \(M_z\sigma_z\) term breaks time-reversal symmetry and induces a mass gap \(\Delta \propto |M_z|\) at \(\mathbf{k}=0\) \cite{qi2011topological}. 

{ In the presence of a finite exchange field $M_z$, time-reversal symmetry is explicitly broken, and the $C_{3v}$ symmetry of the surface is reduced to $C_3$, as the magnetic field breaks mirror symmetry. As a result, the magnetic perturbation not only modifies the
magnitude and spectral profile of the dominant CPGE tensor component
$\sigma_{xxz}$, but also allows additional nonlinear conductivity components,
including $\sigma_{xyz}$ and $\sigma_{xzy}$, to become finite.}
{The relevance of these components can be seen from the helicity-dependent part
of the photocurrent,
\begin{equation}
    \frac{j_x}{{ E}_0^2} = C \sin(2\alpha) + L_1 \sin(4\alpha)  +  L_2 \cos(4\alpha)  +  D ~, \notag   
\end{equation}
where, without any symmetry restrictions, we generally have:
\begin{eqnarray}
C &=&
-i\Bigl[
c_\theta \sigma_{xy}^{-}
-
s_\theta s_\varphi \sigma_{xz}^{-}
+
s_\theta c_\varphi \sigma_{yz}^{-}
\Bigr].
\nonumber
\end{eqnarray}
 Here, $\theta$ and $\varphi$ respectively denote the polar and azimuthal angles, $c/s = \cos/\sin$, and $\sigma_{ab}^{-}$ denotes the antisymmetric, helicity-sensitive part of the nonlinear conductivity tensor. Therefore, the CPGE is controlled not only by the magnitude of individual tensor elements, but also by their antisymmetric combinations. In particular, when $M_z\neq 0$ and the surface symmetry group is reduced to $C_{3}$, $\sigma_{xyz}$ no longer needs to be zero. The relation between $\sigma_{xyz}$ and $\sigma_{xzy}$ then determines whether their contribution enters the circular or linear photocurrent channel (Appendix A). Physically, the emergence of these additional tensor components reflects the redistribution of Berry curvature and the modification of spin textures induced by the exchange field. Thus, the magnetic proximity effect does not simply suppress the original $\sigma_{xxz}$ response by opening a gap; it also redistributes the helicity-dependent response among additional tensor channels.}

\begin{figure}[!t]
    \centering
    \includegraphics[width=0.8\columnwidth]{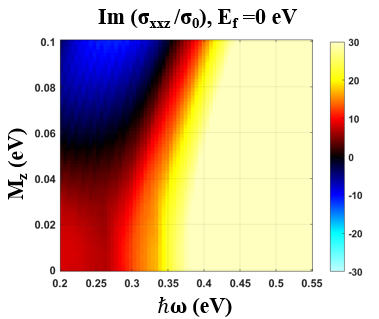}
\caption{Color map of the imaginary part of the conductivity tensor, $\text{Im}(\sigma_{xxz}/\sigma_0)$, versus photon energy $\hbar\omega$ and out-of-plane magnetization $M_z$ (eV) at $E_f = 0$ eV. Colors indicate the sign and magnitude of the response (red/blue for positive/negative), revealing a magnetization-dependent resonance that shifts to higher energies with increasing $M_z$. Note: $\sigma_0 \approx 5.90 \times 10^{-15}~\mathrm{A\,m/V^2}$}

    \label{fig:9}
\end{figure}

\begin{figure}[!t]
    \centering
    \includegraphics[width=0.8\columnwidth]{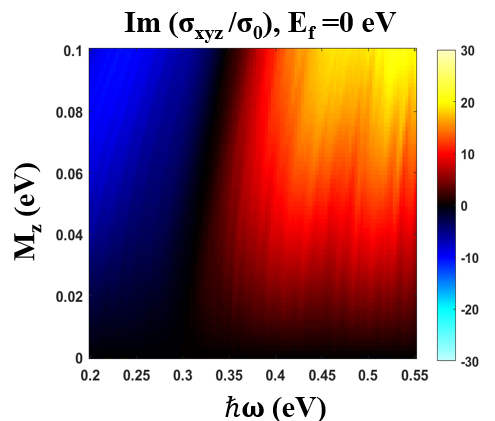}
\caption{Imaginary part of the nonlinear conductivity tensor component $\sigma_{xyz}/\sigma_0$ as a function of photon energy $\hbar\omega$ and magnetization $M_z$, at Fermi energy $E_f = 0~\text{eV}$. The color scale represents the magnitude of $\mathrm{Im}(\sigma_{xyz}/\sigma_0)$, showing a transition from negative (blue) to positive (red) values as $\hbar\omega$ and $M_z$ vary. Note: $\sigma_0 \approx 5.90 \times 10^{-15}~\mathrm{A\,m/V^2}$.}

    \label{fig:10}
\end{figure}

Figure \ref{fig:9} shows the dependence of the top layers' CPGE response, $\Im \sigma^T_{xxz}$, on the exchange field strength $M_z$ at fixed Fermi energy $ E_f = 0$. The $M_z = 0$ line cut is appropriately identical to the $E_f = 0$ line cut in Fig. \ref{fig:5}(a). As $M_z$ is increased from 0, a magnetic gap opens in the surface Dirac spectrum, shifting the onset of interband optical transitions to higher photon energies \(\hbar\omega \approx 2|M_z|\). The  {$\sigma^T_{xxz}$} amplitude near this threshold initially grows due to enhanced asymmetry between spin-split bands, but is subsequently suppressed as the gap widens and available low-energy transitions become Pauli-blocked. The resulting spectrum displays a distinct ridge following the linear relation between the photon energy and \(M_z\), reflecting the magnetic control of the Dirac mass and its role in modulating the helicity-sensitive optical response.
{Figure~\ref{fig:10} illustrates this behavior by showing the imaginary part of $\sigma_{xyz}$ as a function of photon energy $\hbar\omega$ and magnetization $M_z$ at $E_f = 0$. Notably, the color map reveals a continuous redistribution of spectral weight from negative to positive values as the photon energy increases, with a clear magnetization-dependent shift of the response.}\\
In Appendix E, we further explore how for fixed top layer $M_z \neq 0$ the {$\sigma^T_{xxz}$} depends upon the Fermi level location. Figures~\ref{fig:F_2}(a) and (b) show the calculated {$\sigma^T_{xxz}$} response for magnetic configuration Fig. \ref{fig:3}(d) at the top and bottom surfaces of the slab, respectively. They present surface-resolved maps of \(\operatorname{Im}\sigma^{(T/B)}_{xxz}\) with an out-of-plane exchange field \(M_z\). Revisiting the $M_z =0$ case calculated in Fig.\ref{fig:5}(a), a pronounced enhancement appears near the Dirac point at low photon energies, consistent with helicity-selective interband transitions of the gapless, spin–momentum–locked surface states. Introducing $M_z \neq 0$ opens a gap at the Dirac crossing and modifies the spin texture. Two main effects follow: (i) the interband absorption threshold shifts from \(\hbar\omega_{\min} = 0\) to \(\hbar\omega_{\min} \approx 2|M_z|\) for \(E_f \approx 0\) due to Pauli blocking, and (ii) the surface spins acquire finite out-of-plane components, reducing the circular dichroism of the velocity matrix elements. Consequently, the low-energy response is strongly suppressed within the gap window, while spectral weight accumulates just above the band edge. At higher photon energies, the response gradually recovers toward the nonmagnetic profile. In Fig. \ref{fig:F_2}(b), the bottom surface nonlinear conductivity $\Im \sigma^B_{xxz}$ sees little impact of the magnetization in the top layer. It largely retains opposite signs to $\Im \sigma^T_{xxz}$ for the larger photon energies, with small asymmetries arising from finite slab thickness and weak inversion breaking. 

When the exchange field penetrates the entire slab (the “whole-layer” case , Fig. \ref{fig:F_1}(c), it spin–splits the bulk sub bands according to: 
\begin{equation}
E_{n\pm}(k_\parallel) = E_n(k_\parallel) \pm M_z \langle \sigma_z \rangle_n ,
\end{equation} 
lifting Kramers degeneracy and reshaping the joint density of states (JDOS) \cite{litvinov2014magnetic}. The multiple interband thresholds $E_{m+} - E_{n-}$ span a wide energy range, allowing circularly polarized transitions at many photon energies. This produces the broad, smooth CPGE band observed in Fig.~\ref{fig:F_2}(c). The uniform spin polarization also fixes the sign of $\Im \sigma^T_{xxz}$ when $E_f$ lies in the bulk gap; $M_z$ has suppressed directional dependent spin flip transitions.
{ Figure~\ref{fig:E3} further compares the magnetization dependence of
$\mathrm{Im}\,\sigma_{xyz}$ and $\mathrm{Im}\,\sigma_{xzy}$ at fixed photon
energy $\hbar\omega=0.4~\mathrm{eV}$ for different Fermi energies. The two
components evolve differently with $M_z$, producing a finite antisymmetric
combination $\sigma_{xyz}-\sigma_{xzy}$. This antisymmetric part contributes to
the helicity-dependent photocurrent and demonstrates that magnetic proximity
provides an additional route for controlling CPGE beyond the conventional
$\sigma_{xxz}$ channel.}

This behavior highlights an important consequence of magnetic symmetry breaking: the CPGE response is redistributed among multiple tensor components rather than simply suppressed. In particular, the persistence of $\sigma_{xyz}$ demonstrates that helicity-dependent photocurrents can survive inside the magnetically induced gap, providing an additional degree of control over nonlinear optical responses through magnetic tuning.
\section{Conclusion}

We employed a tight-binding slab model and the Kubo formalism to calculate the second-order nonlinear optical conductivity and photocurrent generation in TIs. Our results reveal that both the CPGE and LPGE can be effectively tuned by the Fermi level, gate voltage, photon energy, and surface magnetization. Gate-controlled inversion–symmetry breaking resonantly enhances the nonlinear optical response, particularly in the mid-infrared regime where Berry-curvature effects dominate. Both CPGE and LPGE exhibit strong voltage dependence and complementary spectral features, enabling selective enhancement of polarization-dependent photocurrents.
Surface-state chirality plays a central role: opposite chiralities on the two surfaces lead to partial cancellation of CPGE contributions, whereas same-chirality configurations yield constructive interference and stronger photocurrent generation. This provides a clear strategy for symmetry and spin-texture engineering in polarization-sensitive optoelectronic systems. In addition, proximity-induced magnetization opens a tunable exchange gap in the surface states, redistributing Berry curvature and significantly modulating the CPGE amplitude. Moderate magnetization suppresses the response within the magnetic gap, illustrating the delicate balance between time-reversal-symmetry breaking and optical excitation.

Overall, our findings demonstrate that TIs are a compelling platform for high-frequency, polarization-sensitive, and magnetically controllable, and gate/field-effect–tunable nonlinear optoelectronics. The layered simulation approach presented here enables predictive modeling of how proximate magnetic fields influence nonlinear optical processes at specific surfaces. These results provide theoretical guidance for disentangling intrinsic and extrinsic contributions to CPGE and LPGE observed experimentally and outline design principles for next-generation quantum-geometric and chiral optoelectronic devices operating from the terahertz to mid-infrared spectral range.

\begin{acknowledgments}
This research was partially supported by an Industry University Cooperative Research Center provided by the Army Research Lab (ARL) and in part by the NSF I/UCRC on Multi-functional Integrated System Technology (MIST) Center; IIP-1439644, IIP-1439680, IIP-1738752, IIP-1939009, IIP-1939050, and IIP-1939012. This research was also partially supported by a Laboratory University Collaborative Initiative award provided by the Basic Research Office in the Office of the Under Secretary of Defense for Research and Engineering.
\end{acknowledgments}

\appendix
{\section{Polarization-Dependent Photocurrent}}
{We consider an incident electromagnetic field propagating along the $\hat{r}$ direction with wave vector $\mathbf{k}$. The electric field is transverse to $\mathbf{k}$ and can be decomposed into orthonormal polarization components along the $\hat{\theta}$ and $\hat{\varphi}$ directions. A general polarization state is described by the Jones vector}
{
\begin{equation}
\vec{E}
=
E_\theta \hat{\theta}
+
E_\varphi \hat{\varphi}. \notag
\end{equation}}
{Here, $\theta$ and $\varphi$ denote the polar and azimuthal angles,
respectively, specifying the propagation direction of the incoming light.
The angle $\alpha$ represents the polarization angle, corresponding to
the rotation angle of the quarter-wave plate that controls the polarization
state of the incident radiation. Applying a quarter-wave plate with rotation angle $\alpha$ produces the polarization-dependent electric field}
{\begin{eqnarray}
{\vec{{E}}}
&=&
{{E}}_0
e^{i(qr-\omega t)}
\Big[
(1-i\cos2\alpha)\hat{\theta}
+
i\sin2\alpha\hat{\varphi}
\Big]. \notag
\end{eqnarray}}
{Linear polarization corresponds to $\alpha=0,\pi$, while circular polarization is obtained for $\alpha=\pi/4,3\pi/4$. Using the spherical-coordinate projections}
{\begin{eqnarray}
\hat{\theta}\cdot\hat{x}
&=&
c_\theta c_\varphi,
\qquad
\hat{\theta}\cdot\hat{y}
=
c_\theta s_\varphi,
\qquad
\hat{\theta}\cdot\hat{z}
=
-s_\theta,
\nonumber\\
\hat{\varphi}\cdot\hat{x}
&=&
-s_\varphi,
\qquad
\hat{\varphi}\cdot\hat{y}
=
c_\varphi,
\qquad
\hat{\varphi}\cdot\hat{z}
=
0, \notag
\end{eqnarray}}
{the nonlinear photocurrent along $\hat{x}$ becomes:
\begin{eqnarray}
j_x
&=&
\sigma_{xxx}|{\cal E}_x|^2
+
\sigma_{xxy}{\cal E}_x{\cal E}_y^\ast
+
\sigma_{xyx}{\cal E}_y{\cal E}_x^\ast
\notag\\
&&+
\sigma_{xxz}{\cal E}_x{\cal E}_z^\ast
+
\sigma_{xzx}{\cal E}_z{\cal E}_x^\ast
+
\sigma_{xyy}|{\cal E}_y|^2
\notag \\
&& +
\sigma_{xzz}|{\cal E}_z|^2
+
\sigma_{xyz}{\cal E}_y{\cal E}_z^\ast
+
\sigma_{xzy}{\cal E}_z{\cal E}_y^\ast.
\end{eqnarray}}
{The current can be written compactly as}
{\begin{equation}
\frac{j_x}{{\cal E}_0^2}
=
C\,s_{2\alpha}
+
L_1\,s_{4\alpha}
+
L_2\,c_{4\alpha}
+
D,
\label{eq:A8}
\end{equation}}
{where the coefficient $C$ describes the helicity-dependent circular photocurrent, while $L_1$, $L_2$, and $D$ correspond to linear-polarization contributions. For compactness we define $s = \sin$ and $c = \cos$. It is convenient to define the symmetric and antisymmetric tensor combinations $\sigma_{\alpha\beta}^{\pm}= \sigma_{x\alpha\beta}\pm\sigma_{x\alpha\beta}$. Then we have generally:
\begin{widetext}
\begin{eqnarray}
C
&=&
-i\Bigl[
c_\theta \sigma_{xy}^{-}
-
s_\theta s_\varphi \sigma_{xz}^{-}
+
s_\theta c_\varphi \sigma_{yz}^{-}
\Bigr],
\nonumber\\
L_1
&=&
\frac{1}{2}c_\theta s_{2\varphi}\sigma_{xxx}
-\frac{1}{2}c_\theta c_{2\varphi}\sigma_{xy}^{+}
-\frac{1}{2}s_\theta s_\varphi \sigma_{xz}^{+}
 +\frac{1}{2}s_\theta c_\varphi \sigma_{yz}^{+}
-\frac{1}{2}c_\theta s_{2\varphi}\sigma_{xyy},
\nonumber\\
L_2
&=&
\frac{1}{2}
\left(
c_\theta^2c_\varphi^2-s_\varphi^2
\right)\sigma_{xxx}
+\frac{1}{4}
\left(
1+c_\theta^2
\right)s_{2\varphi}\sigma_{xy}^{+}
-\frac{1}{4}s_{2\theta}c_\varphi\sigma_{xz}^{+}
-\frac{1}{4}s_{2\theta}s_\varphi\sigma_{yz}^{+}
+\frac{1}{2}
\left(
c_\theta^2s_\varphi^2-c_\varphi^2
\right)\sigma_{xyy}
+\frac{1}{2}s_\theta^2\sigma_{xzz},
\nonumber\\[8pt]
D
&=&
\left(
\frac{3}{2}c_\theta^2c_\varphi^2+\frac{1}{2}s_\varphi^2
\right)\sigma_{xxx}
+\frac{1}{4}
\left(
3c_\theta^2-1
\right)s_{2\varphi}\sigma_{xy}^{+}
-\frac{3}{4}s_{2\theta}c_\varphi\sigma_{xz}^{+}
-\frac{3}{4}s_{2\theta}s_\varphi\sigma_{yz}^{+}
+\left(
\frac{3}{2}c_\theta^2s_\varphi^2+\frac{1}{2}c_\varphi^2
\right)\sigma_{xyy}
+\frac{3}{2}s_\theta^2\sigma_{xzz}.\notag
\end{eqnarray}
\end{widetext}
For circular polarization, $\alpha=\pi/4,3\pi/4$, and so one has $s_{2\alpha}=\pm1$, $ s_{4\alpha}=0$, $c_{4\alpha}=-1$, so that only the coefficient $C$ changes sign with helicity. The CPGE response is therefore controlled by the antisymmetric conductivity components $\sigma_{\alpha\beta}^-$. In contrast, for linear polarization, $ \alpha=0,\pi$, the helicity-dependent term vanishes and the photocurrent depends only on the symmetric tensor components.}

{The nonlinear conductivity tensor vanishes in inversion-symmetric systems because the current changes sign under spatial inversion while the quadratic electric-field term remains invariant.  Consequently, inversion-symmetry breaking mechanisms such as surface potentials, finite optical penetration depth, gate-induced potentials, or magnetic proximity effects are required to generate a finite CPGE response in topological-insulator slabs. If we limit ourselves to the discussion of the Bi$_2$Se$_3$ surface, which possesses $C_{3v}$ symmetry, it has been shown the photogalvanic coefficients reduce to \cite{connelly2024emergence}:
\begin{eqnarray}
    C &=& -i[c_\theta \sigma^-_{xy} - s_\theta s_\varphi \sigma^-_{xz}]\nonumber\\
    L_1 &=& \dfrac{\sigma_{xxx}}{2}c_\theta s_{2\varphi} -\dfrac{1}{2}[c_\theta c_{2\varphi}\sigma^+_{xy}+s_\theta s_\varphi\sigma^+_{xz}]\nt\\
    &&-\dfrac{1}{2}c_\theta s_{2\varphi}{\sigma_{xyy}}\nonumber\\
    L_2 &=& \dfrac{\sigma_{xxx}}{2}(c_\theta^2c_\varphi^2) -\dfrac{1}{4}\left[\left(c_\theta^2 s_{2\varphi}\right)\sigma^+_{xy}+ s_{2\theta}c_\varphi\sigma^+_{xz}\right]\nt\\
    && +s_\theta^2\sigma_{xzz}+\dfrac{1}{2}c_\theta^2 s_{\varphi}^2\sigma_{xyy}\nonumber
\end{eqnarray}}
{Time-reversal symmetry breaking is not necessary for a nonzero CPGE, but magnetic order can strongly enhance the response by redistributing the Berry curvature near the gap edge and increasing helicity-selective interband transitions.

{\section{Injection and Shift Current Contributions to LPGE and CPGE}
\renewcommand{\thefigure}{B.\arabic{figure}}
\setcounter{figure}{0}

\begin{figure}[b] % placement: t=top, b=bottom, h=here, H=require exact (needs float)
  \centering
    \includegraphics[width=0.9\columnwidth]{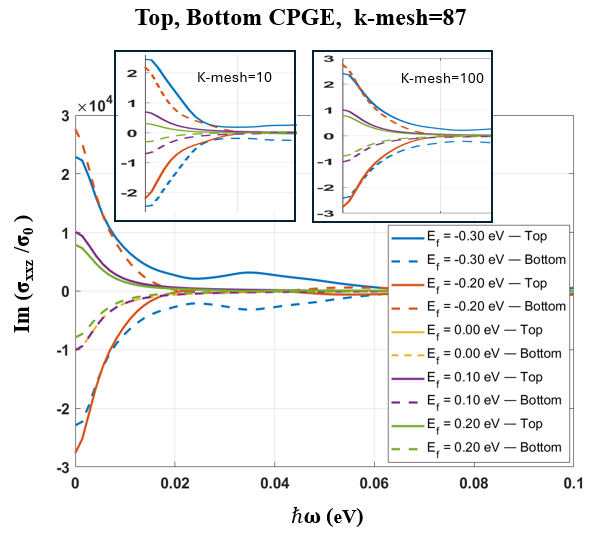}%
 % or .png/.jpg
 \caption{Imaginary part of the nonlinear conductivity $\sigma_{xxz}$ for the
  top (solid) and bottom (dashed) surfaces of the slab versus photon
  energy $\hbar\omega$. Colors pair top/bottom curves at the same Fermi level
  $E_f\in\{-0.2, 0.0, 0.2\} $eV. Over most of the spectrum the two surfaces
  contribute with nearly equal magnitude and opposite sign, consistent with the
  odd parity of $\sigma_{xxz}$ under $z\!\to\!-z$. Note: $\sigma_0  \approx  6.85\times 10^{-7}  \text{A}/\text{V}^2$.}
  \label{fig:D1}
\end{figure}

In nonlinear optical response theory, the second-order conductivity tensor contains distinct microscopic contributions associated with injection and shift currents.  The injection current originates from asymmetric carrier population dynamics under optical excitation, while the shift current arises from the real-space displacement of Bloch electrons during interband transitions. 
Starting from the general second-order conductivity tensor, these two contributions can be separated naturally according to their resonant structure and geometric properties.}
\begin{widetext}
{The nonlinear optical conductivity is given by:}
{\begin{equation}
\begin{split}
\sigma_{\mu \alpha \beta}(\omega; \omega_1, \omega_2)
&=
\frac{-e^3}{\hbar^2 \omega_1 \omega_2}
\sum_{a,b,c}
\int [d\mathbf{k}]
\,
\Bigg[
\frac{1}{2}
f_a
\, h^{\mu \alpha \beta}_{aa}
\quad+
f_{ab}
\frac{
v^\alpha_{ab}
h^{\mu \beta}_{ba}
}{
\omega_1 - \varepsilon_{ab}
}
+
f_{ab}
\frac{
\tfrac{1}{2}
h^{\alpha \beta}_{ab}
v^\mu_{ba}
}{
\omega - \varepsilon_{ab}
}
\\
&\quad+
v^\alpha_{ab}
v^\beta_{bc}
v^\mu_{ca}
\frac{
(\omega_2 - \varepsilon_{cb}) f_{ab}
+
(\omega_1 - \varepsilon_{ba}) f_{cb}
}{
(\omega_1 - \varepsilon_{ba})
(\omega_2 - \varepsilon_{cb})
(\omega - \varepsilon_{ca})
}
\quad+
[(\alpha,\omega_1)\leftrightarrow(\beta,\omega_2)]
\Bigg].
\label{eq:A6}
\end{split}
\end{equation}}
\end{widetext}
{The first {diagonally indexed} term is the {intraband} Fermi surface response, the second and third comprise the {two-band interband} shift current response, and the {fourth} is the {three-band resonant } injection {current}. The terms $h^{\mu \alpha}$ and $h^{\mu \alpha \beta}$ are higher order covariant derivatives of the velocity operator:
\begin{equation}
h^{\mu \beta} =(D^\mu V^\beta)_{ba} = \partial_{k_\mu}V^\beta_{ba}
-
i
\big(
\mathcal A^\mu_{bb}
-
\mathcal A^\mu_{aa}
\big)
V^\beta_{ba},
\end{equation}}

We can simplify the shift current into the form:
\begin{eqnarray}
\sigma_{\mu\alpha\beta}^{\rm shift}(\omega;\omega_1,\omega_2)
=
-\frac{e^3}{\hbar^2\omega_1\omega_2} \text{Tr} \left[T_1^{\mu \alpha \beta} + T_2^{\mu \alpha \beta} + \frac{1}{2} T_3^{\mu \alpha \beta} \right]~,\nt
\end{eqnarray}
where we have defined the terms:
\begin{gather}
(T_1^{\mu\alpha\beta})_{ij} = \frac{f_{ij} V^\alpha_{ij}h^{\mu\beta}_{ji}}{\omega_1 + \varepsilon_{ij}}~, ~~ (T_2^{\mu\alpha\beta})_{ij}= \frac{f_{ij} V^\beta_{ij}h^{\mu\alpha}_{ji}}{\omega_2 + \varepsilon_{ij}}~,\notag \\
(T_3^{\mu\alpha\beta})_{ij}= \frac{f_{ij} \left(h^{\alpha\beta}_{ij}+h^{\beta\alpha}_{ij}\right)V^\mu_{ji}}{\Omega + \varepsilon_{ij}}~.\nt
\end{gather}
In order to avoid numerical gauge issues from taking outer derivatives of inner products (i.e. $\del_{k_\alpha} \langle n|H(\mathbf{k})|m\rangle$) we have expressed the covariant derivative of velocity in terms of operators:
\begin{eqnarray}
h^{\alpha\beta}_{ij}
&=&
(D^\alpha v^\beta)_{ij} = \partial_{k_\alpha} v^\beta_{ij}
-
i(\mathcal A^\alpha_{ii}-\mathcal A^\alpha_{jj})v^\beta_{ij} \notag\\
&=&
\frac{1}{\hbar}\langle i|\partial_{k_\alpha}\partial_{k_\beta}H|j\rangle
+ \left[\frac{V^\alpha}{\tilde{\varepsilon}},V^\beta \right]_{ij}~.\nt
\end{eqnarray}
In the main text we have shown that the injection current can be expressed as: 
\begin{eqnarray}
\sigma_{\mu \alpha \beta }^{\text{inj}} =  \frac{e^3/\hbar^2}{\omega_1 \omega_2}   \text{Tr} \left[ t_0^{\mu} \left( t_1^{\alpha} V^{\beta} + t_2^{\beta} V^{\alpha} - V^{\alpha} t_2^{\beta} - V^{\beta} t_1^{\alpha} \right) \right]~.\nt
%\label{injsimp}
\end{eqnarray}
This can be further simplified to the standard injection current formula from the literature \cite{xie2025photon}. Consider the DC limit of $\omega_1 = \omega + i \delta$, $\omega_2 = -\omega + i\delta$, in which case $1/(\Omega + 2 i\delta - \varepsilon_{ca}) \approx \delta_{ca}/(2 i \delta) - (1-\delta_{ca})/ \varepsilon_{ca}$. The leading resonant contribution comes from the $\delta_{ac}/(2 i \delta)$ term. When $c =a$, the $t^\alpha_1$ and $t^\beta_2$ frequency dependent denominators can be combined into $\pi i \delta(\varepsilon_{ba}-\omega)$. Cleaning up the resulting expression we obtain the familiar result:
\begin{eqnarray}
    \sigma_{\mu \alpha \beta }^{\text{inj}} &=& \frac{\pi e^3 /\hbar^2}{2 \delta } \frac{1}{\omega^2} \sum_{i,j,\mathbf{k}} (V^\mu_{ii}-V^\mu_{jj})V^\alpha_{ij}V^\beta_{ji} f_{ij} \delta(\varepsilon_{ij}-\omega) \nt\\
    &&+ \mathcal{O}\left({\frac{1}{E_g}}\right)~,
\end{eqnarray}
where we have put the off-resonant terms into the $\mathcal{O}(1/E_g)$ term that is suppressed by the band gap. {
We can now estimate the relative magnitude of the contributions of shift and injection currents to the CPGE and LPGE. Following our discussion in Appendix A, for three-fold symmetric Bi$_2$Se$_3$ this is primarily encoded in the ratio:
\begin{equation}
\frac{-i (\sigma_{\mu \alpha \beta}-\sigma_{\mu \beta \alpha})}
{(\sigma_{\mu \alpha \beta}+\sigma_{\mu \beta \alpha})}\equiv \frac{\mathrm{Im} ~\sigma_{\mu\alpha\beta}}
{\mathrm{Re}~ \sigma_{\mu\alpha\beta}}\equiv \frac{\text{CPGE}}{\text{LPGE}}~,
\end{equation}
where $\mathrm{Im}(\sigma_{xxz})$ corresponds to the CPGE contribution, while $\mathrm{Re}(\sigma_{xxy})$ describes the LPGE response in the chosen geometry. The goal is to determine which nonlinear response dominates under near-resonant optical excitation and to clarify how the competition between CPGE and LPGE depends on resonance detuning, broadening, and matrix-element effects.}
\subsection{Shift Ratio}
In the DC limit notice in $\sigma_{\mu\alpha\beta}^{\rm shift}(\omega;\omega_1,\omega_2)$ that $T_1^{\mu \alpha \beta}(\omega_1) = T_2^{\mu\beta\alpha}(\omega_1) $ and $T_3^{\mu \alpha \beta} = T_3^{\mu\beta\alpha}$. Then the contribution to CPGE from the shift current goes as:
\begin{eqnarray}
    i(\sigma_{\mu\alpha\beta}^{\text{shift}}-\sigma_{\mu\beta\alpha}^{\text{shift}}) \sim \frac{1}{\omega^2}\sum \frac{\omega f_{ij}}{\varepsilon_{ij}^2+\omega^2}\left( V^\alpha_{ij} h^{\mu\beta}_{ji} - V^\beta_{ij} h^{\mu\alpha}_{ji}  \right)~, \nt
\end{eqnarray}
while the LPGE contribution is:
\begin{eqnarray}
    (\sigma_{\mu\alpha\beta}^{\text{shift}}+\sigma_{\mu\beta\alpha}^{\text{shift}}) &\sim& \frac{1}{\omega^2}\sum \frac{\varepsilon_{ij} f_{ij}}{\varepsilon_{ij}^2+\omega^2}\left( V^\alpha_{ij} h^{\mu \beta}_{ji} + V^\beta_{ij} h^{\mu\alpha}_{ji}  \right) \notag\\
    &~& + \frac{1}{\omega^2} \Tr T^{\mu\alpha\beta}_3 ~. \nt
\end{eqnarray}
The Fermi-Dirac different $f_{ij}$ in $T_3$ eliminates the $i = j$ sums, and thus there is no $1/\delta$ contribution as in the injection current, so $T^3 \sim \mathcal{O}(1/E_g)$. The non-vanishing $T^3$ contribution guarantees that in general LPGE$^{\text{shift}}$ $>$ CPGE$^{\text{shift}}$.

\subsection{Injection Ratio}
We have shown that to first order the injection current scales as $1/\delta$. Then naively one would expect CPGE$^\text{inj}$ $\sim$ LPGE$^{\text{inj}}$, but we see from the results in the main text that this is not the case. While explicit calculation is the clearest way to determine the relative contributions, we can motivate the smaller LPGE contribution by observing $(V^{\alpha}_{ij})^* = V^{\alpha}_{ji}$, from which follows:
\begin{eqnarray}
    i(\sigma_{\mu\alpha\beta}^{\text{inj}}-\sigma_{\mu\beta\alpha}^{\text{inj}}) &\sim& \frac{1}{\delta \omega^2} \sum_{i,j,\mathbf{k}} \mathcal{A}_{ij}^\mu \text{Im}\left(V^\alpha_{ij}V^\beta_{ji}\right)  + \mathcal{O}\left({\frac{1}{E_g}}\right)~, \nt\\
   (\sigma_{\mu\alpha\beta}^{\text{inj}}+\sigma_{\mu\beta\alpha}^{\text{inj}}) &\sim& \frac{1}{\delta \omega^2} \sum_{i,j,\mathbf{k}} \mathcal{A}_{ij}^\mu \text{Re}\left(V^\alpha_{ij}V^\beta_{ji}\right)  + \mathcal{O}\left({\frac{1}{E_g}}\right)~.\nt
\end{eqnarray}
For simple surface state models, it has previously been shown that the term Re$(V^{x}_{ij}V^{y}_{ji})$ will lead to a net-zero $\mathbf{k}$ integration if there is no warping present in the band structure \cite{Li2019,han2024design,leppenen2023linear}. Given the weak warping of Bi$_2$Se$_3$ near the Dirac point (Fig. 1), the $\mathcal{O}(1/\delta)$ leading contribution of the injection current is overall suppressed in  LPGE, and so the next leading order comes from the $\mathcal{O}(1/E_g)$ term. The CPGE to LPGE ratio for $\hbar \omega \lesssim 1$ eV is thus given by:
\begin{eqnarray}
    \frac{\text{CPGE}^{\text{inj}}}{\text{LPGE}^{\text{inj}}} \sim \frac{E_g}{\delta} \sim 100~,
\end{eqnarray}
which is consistent with the results presented in Fig. \ref{fig:4}. At higher energies, both CPGE and LPGE decrease as the joint density of states drops off. By imposing stronger warping, using a model we discuss in the following section, the injection contribution to LPGE can be greatly enhanced. 

Finally, we note smaller values of $\delta$ require k-mesh densities that push past our computational memory limitations. In Fig. \ref{fig:D1}, in addition to the antisymmetric response of the top/bottom TI surface states, we demonstrate the k-mesh dependent numerical convergence of the low frequency behavior of the injection current for $\delta = 0.005$ eV, which reflects a clean, long quasi particle lifetime. 
}
\section{Model Parameters and Warping Choice}
\renewcommand{\thefigure}{C.\arabic{figure}}
\setcounter{figure}{0}
\begin{figure}[!t]
    \centering
    % --- Left Figure ---
    \begin{minipage}{0.48\textwidth}
        \centering
        \includegraphics[width=\linewidth]{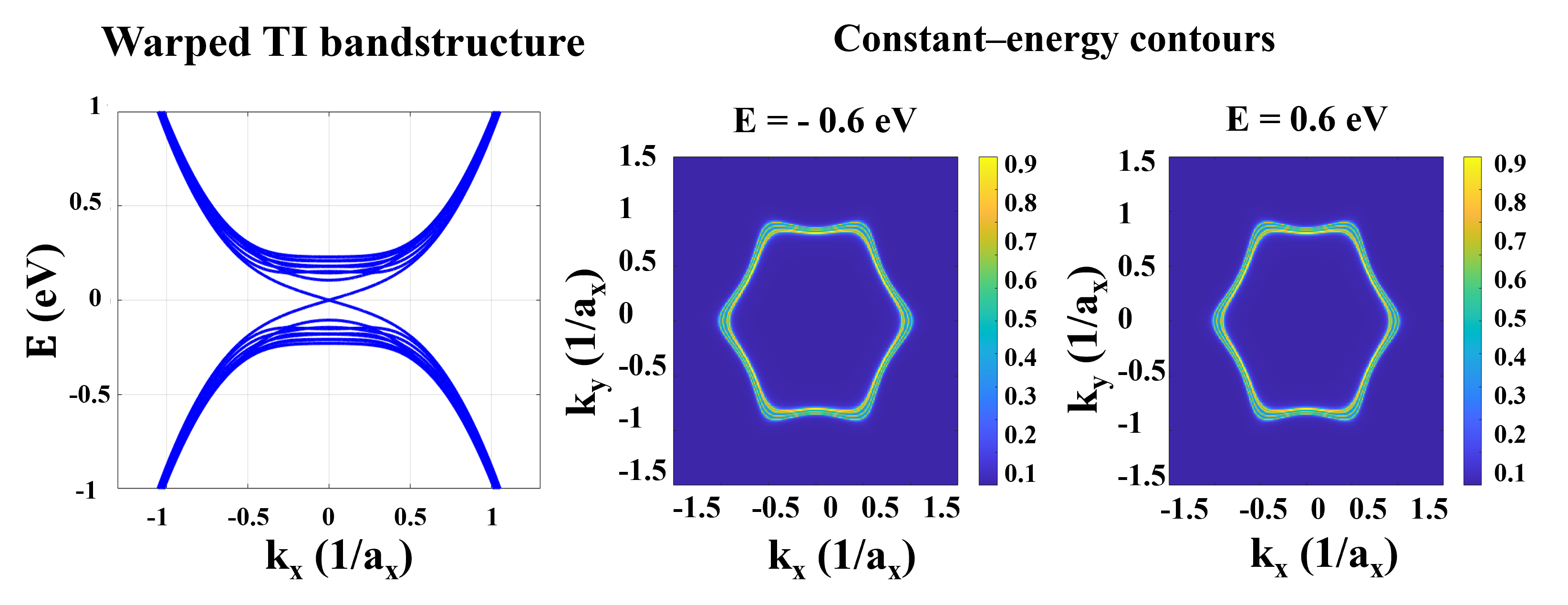}
        \caption{Electronic structure of warped TI: (Left) Modification of the Bi$_2$Se$_3$ parameters to engineer greater warping in the band structure. The constant-energy contours are seen to deform into hexagon/snowflake shapes in conduction band.}
        \label{fig:B_2}
    \end{minipage}
    \hfill % Adds flexible space between the two images
    % --- Right Figure ---
    \begin{minipage}{0.48\textwidth}
        \centering
        \includegraphics[width=\linewidth]{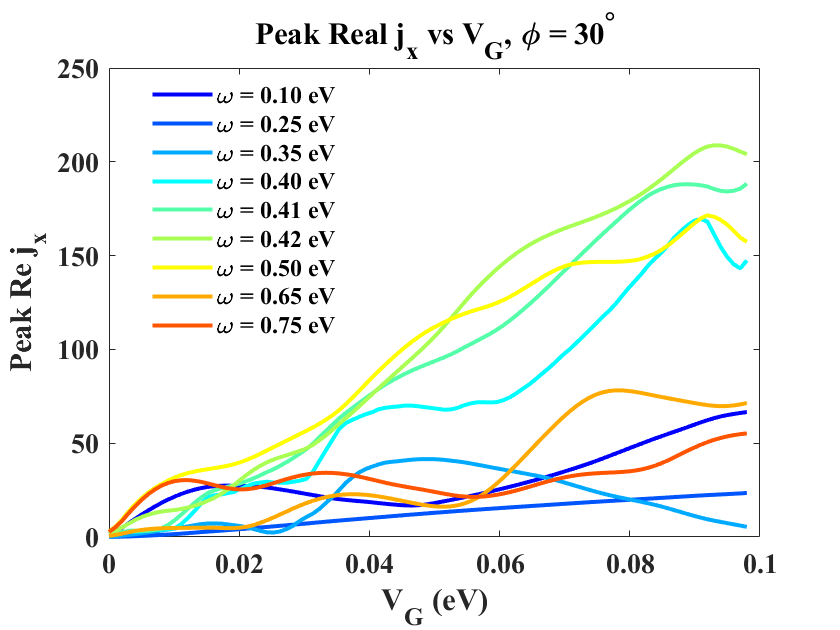}
        \caption{Generated photocurrent component $j_{x}$ as a function of applied field strength for different photon energies $\omega$, with the incident angle fixed at $\varphi = 30^{\circ}$. The results are obtained from the nonlinear optical response of the TI slab model, incorporating hexagonal warping effects.}
        \label{fig:B_3}
    \end{minipage}
\end{figure}

For the four and eight band tight binding models used in the main text we use lattice constants $a=4.14~\text{\AA}$ and $c=28.64~\text{\AA}$. The nine real parameters that comprise the model are collected as
\begin{equation}
\mathbf{A} = [m_{11},~A_0,~A_{11},~A_{14},~A_{12},~B_0,~B_{11},~B_{14},~B_{12}]~.
\notag
\end{equation}
Their impact can be summarized as:
\begin{itemize}[leftmargin=*]
  \item $m_{11}$: Band inversion or mass term (adds to $\Gamma_5$).
  \item $A_0$: Isotropic in-plane scalar dispersion (background curvature).
  \item $A_{11}$: Even-in-$k$ term coupling to $\Gamma_5$, shifts the bulk gap.
  \item $A_{14}$: In-plane anisotropic (odd-in-$k$) term with trigonal structure; drives hexagonal warping via $\Gamma_{1,2}$.
  \item $A_{12}$: Isotropic odd-in-$k$ term (couples to $\Gamma_3$); controls Dirac velocity and contributes to warping.
  \item $B_0$: Scalar interlayer hopping amplitude.
  \item $B_{11}$: Even interlayer term coupled to $\Gamma_5$.
  \item $B_{14}$: Anisotropic interlayer term coupled to $\Gamma_{1,2}$ (trigonal structure).
  \item $B_{12}$: Isotropic interlayer term coupled to $\Gamma_4$.
\end{itemize}
We set the numerical parameters in units of eV as:\\
(i) Baseline Bi$_2$Se$_3$:
\begin{eqnarray}
\mathbf{A}_{\mathrm{Bi_2Se_3}} &=&
[10.2,-1.175,-1.72,0.369,-2.85, \notag\\
&&-0.021,0.075,-0.175,-0.039]. \notag
\end{eqnarray}
(ii) ``Highly Warped'' (Enhanced Hexagonal Warping):
\begin{eqnarray}
\mathbf{A}_{\mathrm{warp}} &=& [ 2.55, 0, -0.43, 0.092, -2.85, \notag \\
&& 0, 0.019, -0.044, -0.039 ] \notag
\end{eqnarray}
The ``warped'' set suppresses the isotropic background curvature and interlayer scalar hopping by setting $A_0=B_0=0$. 
The remaining coefficients $m_{11}, A_{11}, A_{14}, B_{11}, B_{14}$ are scaled by a factor of $1/4$ to reduce the overall bandwidth and amplify the relative effect of the anisotropic odd-in-$k$ terms ($A_{14}$ and $B_{14}$). 
This adjustment increases the visibility of hexagonal warping in constant-energy contours while preserving the low-energy Dirac structure through unchanged $A_{12}$ and $B_{12}$.

Figure \ref{fig:B_3} presents the evolution of the maximum real part of the photogenerated current density, $j_x$, as a function of the applied gate voltage, evaluated at a fixed angle of incidence $30^{\circ}$ for various photon energies. The functional form of $j_x$ is derived in Appendix A. The results in Fig. \ref{fig:5} demonstrate that the nonlinear photogalvanic response is highly sensitive to both the excitation frequency and the external voltage. At lower photon energies (0.1--0.4~eV), the generated current remains relatively small and exhibits weak dependence on the applied voltage, indicating that the system operates within the perturbative regime with minimal nonlinear amplification. In contrast, at higher photon energies, particularly near $\omega \approx 0.5$~eV, the photogenerated current increases sharply with increasing voltage. This strong enhancement is attributed to resonant interband transitions, wherein the incident photon energy matches the $\sim$ 0.5 eV bulk energy band gap of the model. These resonances enhance the nonlinear optical conductivity components responsible for both CPGE and LPGE, leading to a rapid increase in the total photocurrent.

\section{Responsivity}
In this Appendix we calculate the responsivity of our device following the data in Fig. \ref{fig:6}. The responsivity is defined as the ratio of photocurrent to incoming optical power:
\begin{equation}
R = \frac{I_{\mathrm{photo}}}{P_{\mathrm{in}}}
\end{equation}
The photocurrent we calculate  originates from the first two layers of the TI, and would be generated by a mid-wave infrared laser of typical spot size diameter 250 $\mu$m. Therefore
\begin{eqnarray}
    I_{\text{photo}} &=& J_{calc}\times (2~\text{nm}) \times (250~\mu\text{m}) \nt\\
    &=& \sigma_{xxz} E^2 \times (2~\text{nm}) \times (250~\mu\text{m})
\end{eqnarray}
The electric field strength is related to the intensity of the laser, $i_l$ by 
\begin{gather}
i_l = \frac{1}{2}c\varepsilon_0 E^2
  \Longrightarrow  
E^2 = \frac{2 i_l }{c\varepsilon_0}~,\nt\\
c\varepsilon_0 \approx 2.655 ~\text{mS} \ \Rightarrow\ 
E^2 \approx (753 \Omega)  i_l~.\nt
\end{gather}
The power incident, $P_{\text{in}}$ on a spot size diameter 250 $\mu$m is given by:
\begin{equation}
    P_{\text{in}} = i_l \times \pi (125~\mu\text{m})^2~.
\end{equation}
Putting everything together we have:
\begin{eqnarray}
R &=& \frac{I_{\mathrm{photo}}}{P_{\mathrm{in}}} = \frac{(753 \Omega) i_l \sigma_{xxz} \times (2~\text{nm}) \times (250~\mu\text{m}) }{i_l \times \pi (125~\mu\text{m})^2} \nt\\
&=& (7.67~\text{m}\Omega) \sigma_{xxz}
\end{eqnarray}
For the CPGE at $\hbar \omega = 0.4$ eV we extract $\Im \sigma_{xxz}^T = 2.21\times 10^{-5}$ A$/$V$^2$, and finally $R = 1.695 \times 10^{-7}$ A$/$W.

\section{Magnetization}
\renewcommand{\thefigure}{E.\arabic{figure}}
\setcounter{figure}{0}
{The presence of magnetization gaps the surface state as shown in Fig. \ref{fig:F_1}. We can see that a local magnetic field will only gap one surface for our Hamiltonian. The resulting implications for the $\sigma_{xxz}$ and $\sigma_{xyz}$ tensors are presented in Figs. \ref{fig:F_2} and \ref{fig:E3}.}

\begin{figure}[!h]
    \centering
    \includegraphics[width=\columnwidth]{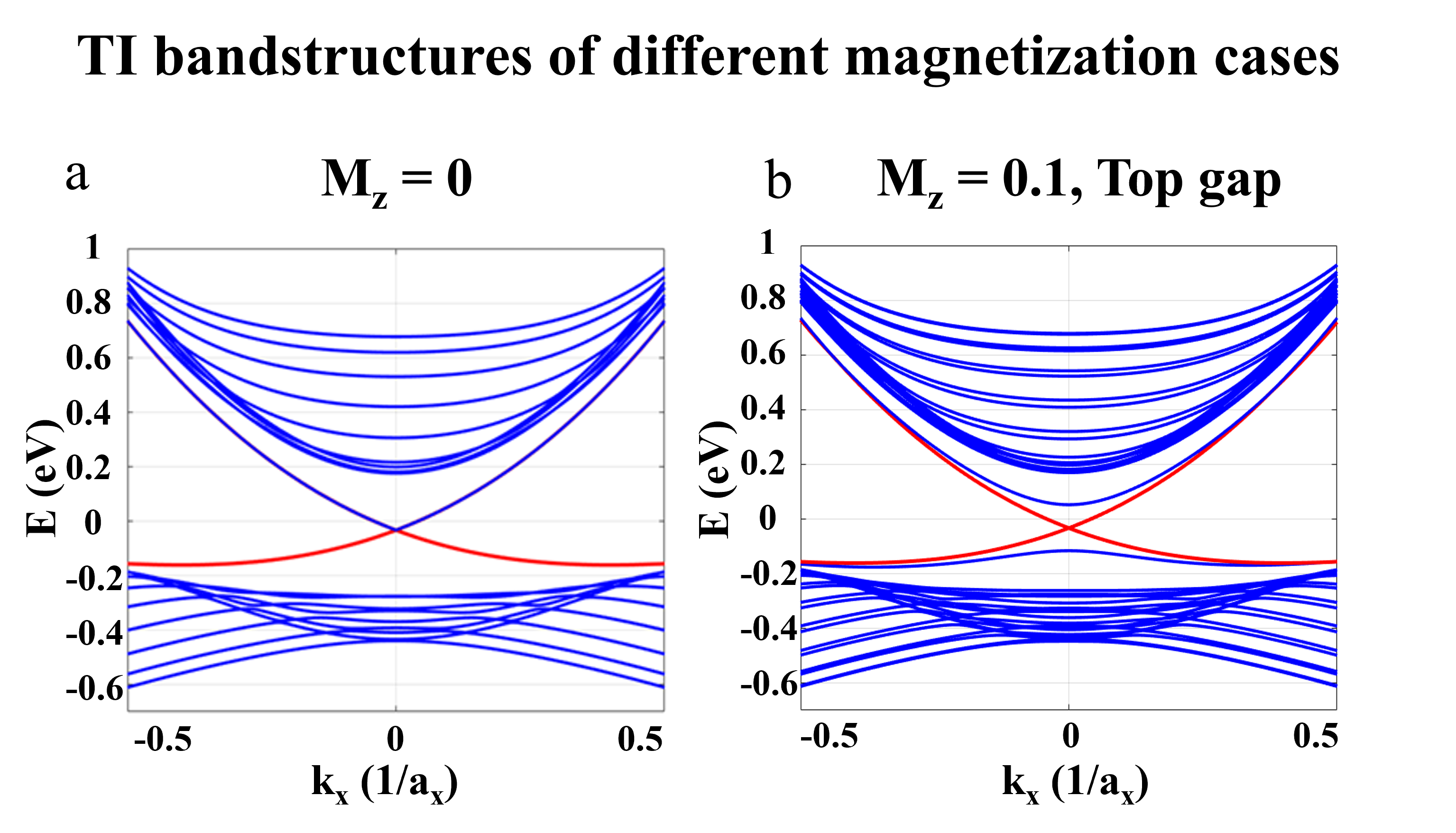}
     \centering
     \includegraphics[width=0.55\columnwidth]{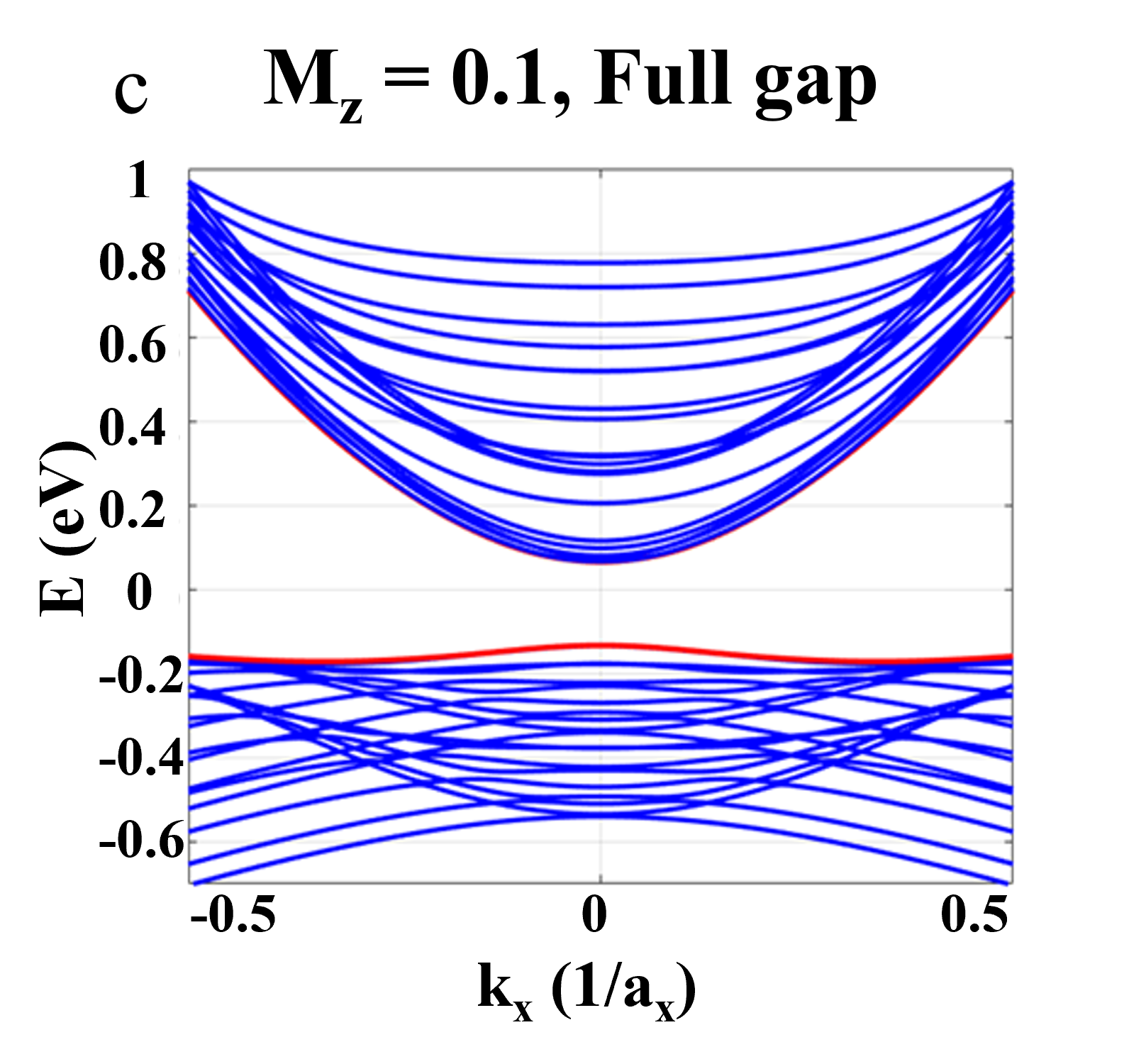}
\caption{Band structure of a TI slab for different out-of-plane exchange field strengths. 
a) $M_z=0$, showing gapless Dirac surface states at the $\Gamma$ point. 
b) surface magnetization. $M_z=0.1$, where the exchange field opens a gap in the surface states. c) bulk magnetization. $M_z=0.1$, where the exchange field opens a full gap.
Bulk bands are shown in blue, and surface states are highlighted in red.}
    \label{fig:F_1}
\end{figure}
\begin{figure}[!h] % placement: t=top, b=bottom, h=here, H=require exact (needs float)
  \centering
  \includegraphics[width=0.9\columnwidth]{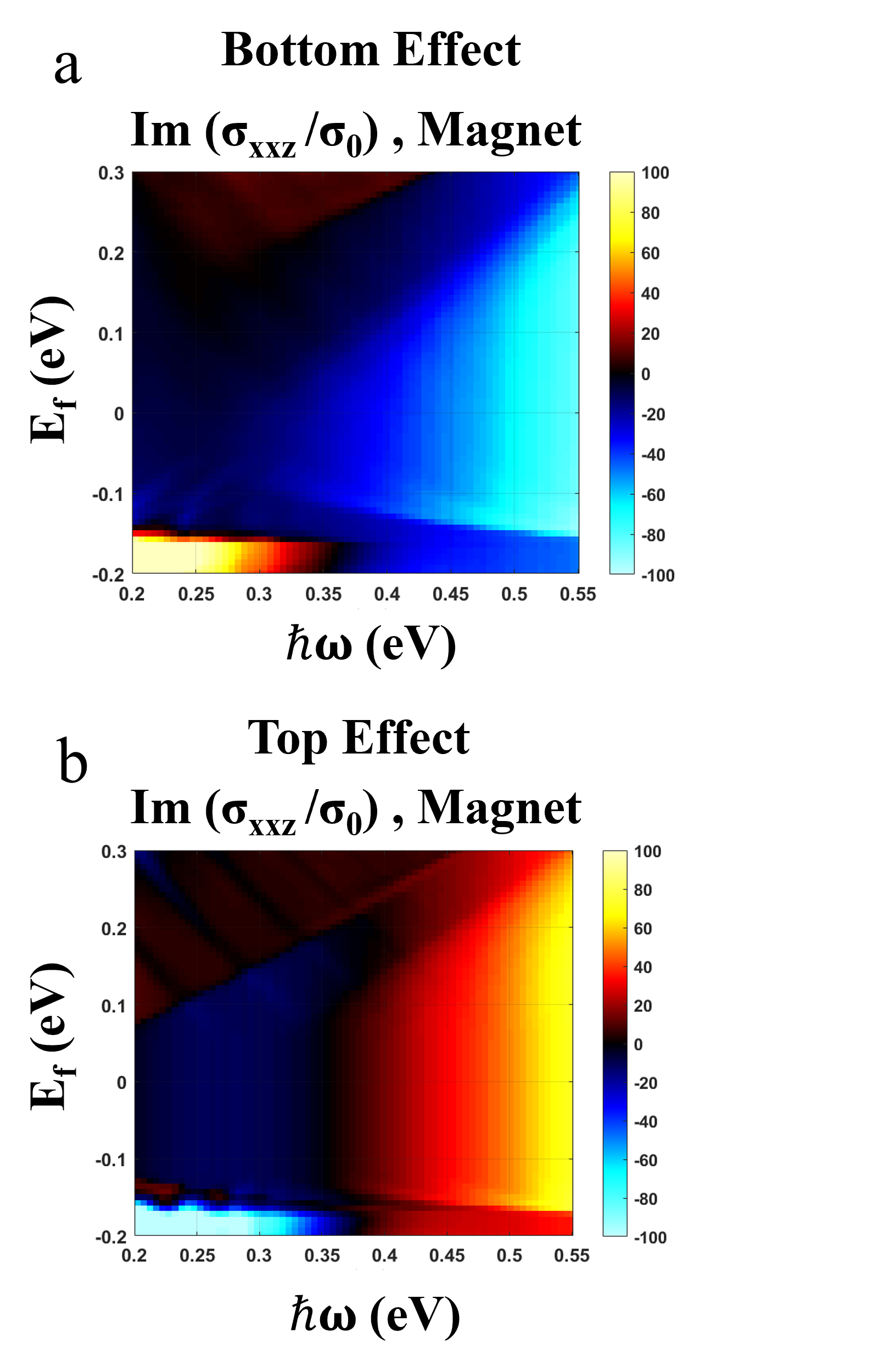}
    \includegraphics[width=0.9\columnwidth]{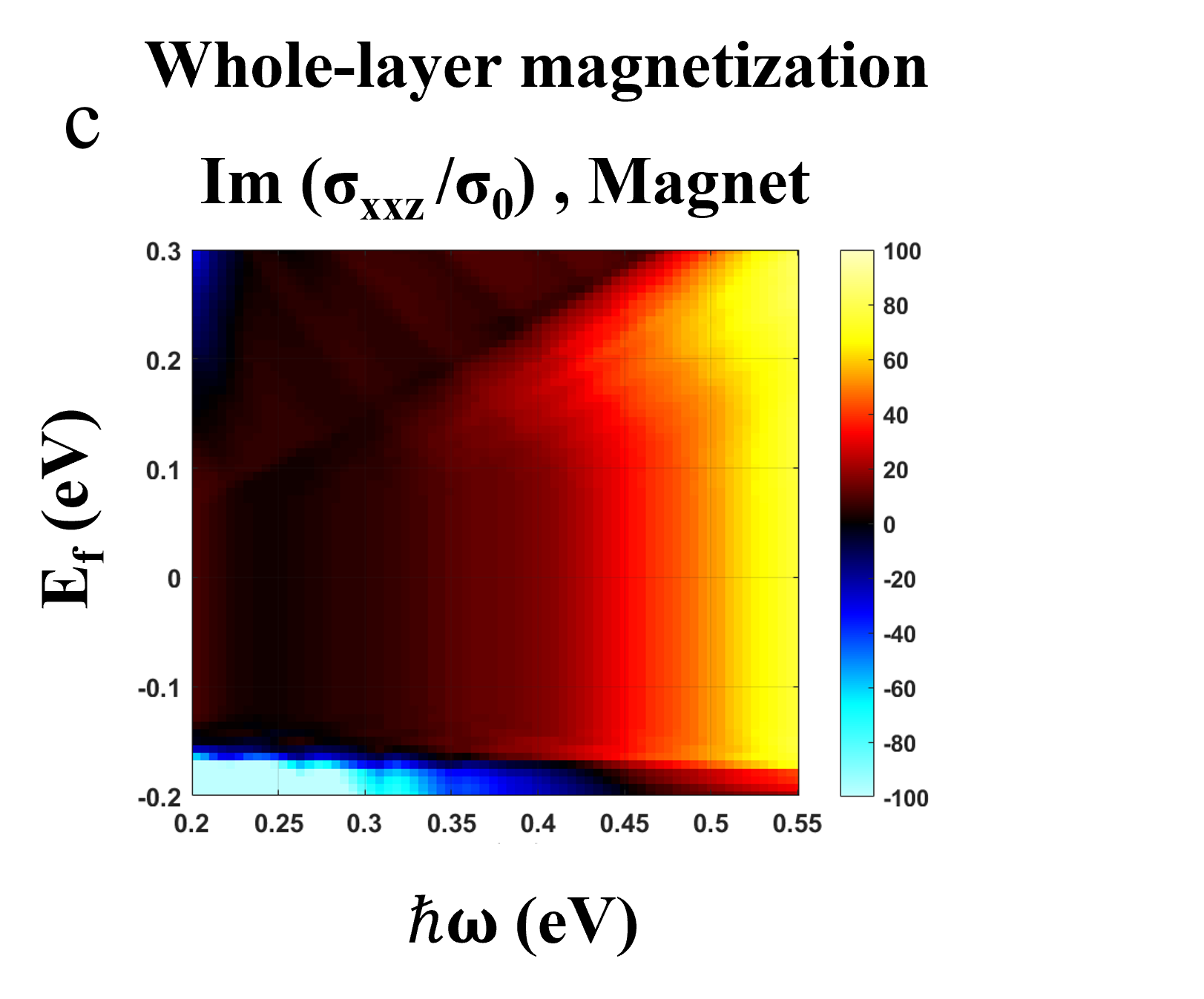}
   \caption{(a) and (b) Imaginary part of the second-order optical conductivity component $\text{Im}(\sigma_{xxz})$ as a function of photon energy $\hbar\omega$ and Fermi level $E_f$, at bottom and top respectively. The magnetic exchange gap generated by $M_z = 0.1$ eV suppresses CPGE within the gap region. Whether the magnet is collocated with the surface state (Top) or on the opposite surface (bottom), its influence persists, suppressing CPGE and potentially reversing its sign. c) for magnetization $M_z = 0.1$ eV through the slab as in Fig. \ref{fig:F_1}(c). Extending the magnetization across the layers make the response stronger, while the overall magnitude and sign pattern are preserved.  $\sigma_0 \approx 5.90 \times 10^{-15}~\mathrm{A\,m/V^2}$.}
    \label{fig:F_2}
\end{figure}

\begin{figure}[t!] % placement: t=top, b=bottom, h=here, H=require exact (needs float)
  \centering
    \includegraphics[width=\columnwidth]{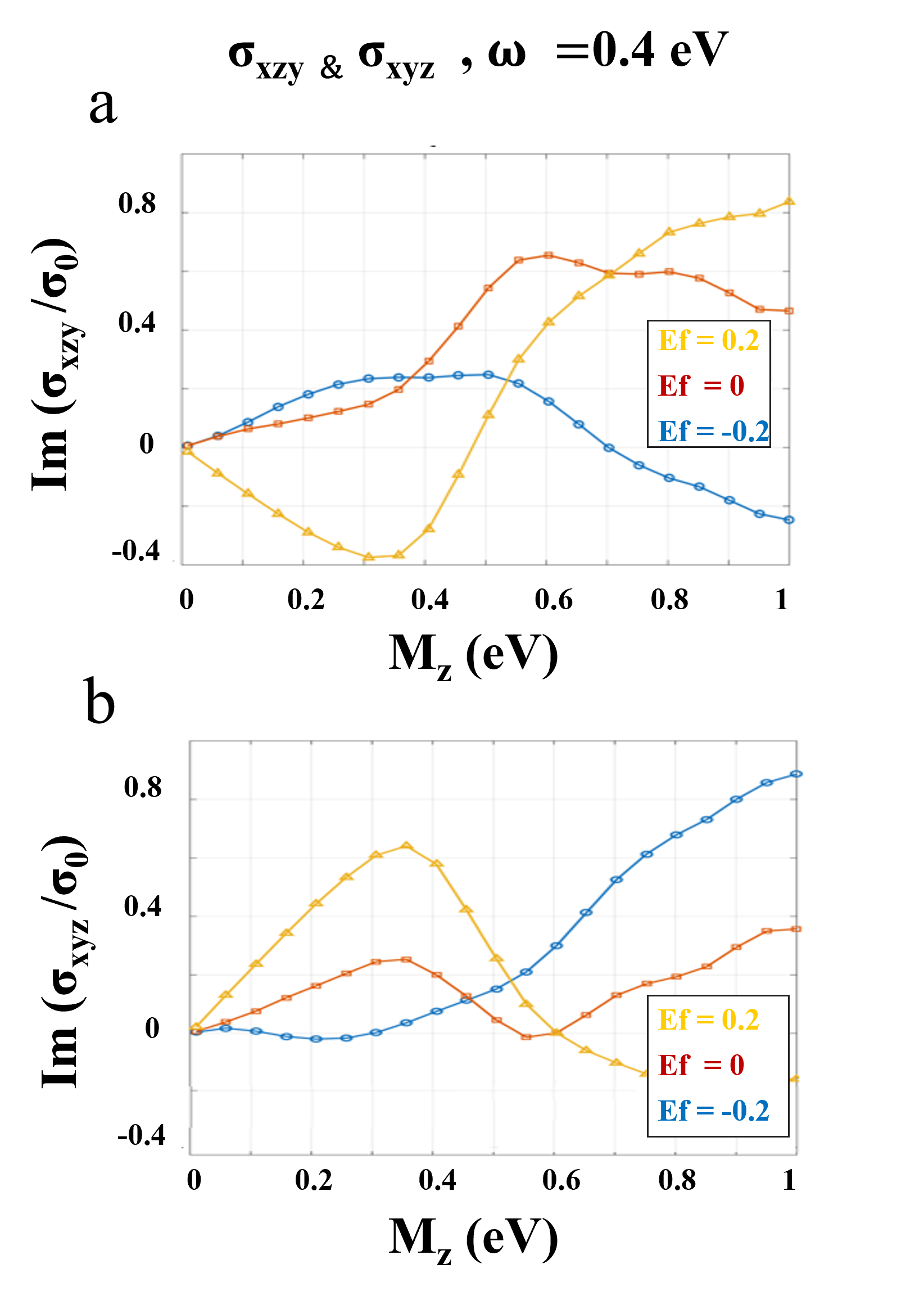}%
 % or .png/.jpg
\caption{
Magnetization dependence of the imaginary part of the nonlinear optical conductivity tensor components at fixed photon energy $\hbar\omega = 0.4~\mathrm{eV}$. 
a) $\mathrm{Im}(\sigma_{xzy}/\sigma_0)$; b) $\mathrm{Im}(\sigma_{xyz}/\sigma_0)$, plotted as functions of out-of-plane magnetization $M_z$ for different Fermi energies $E_f = -0.2,\,0,\,0.2~\mathrm{eV}$. 
The results illustrate the emergence and evolution of magnetization-induced nonlinear responses, including sign changes and enhancement of tensor components due to time-reversal symmetry breaking and the redistribution of Berry curvature. 
In particular, the finite difference $\sigma_{xyz} - \sigma_{xzy} \neq 0$ highlights the appearance of an antisymmetric contribution, indicating that the CPGE acquires an additional term in the presence of magnetization. 
Note: $\sigma_0 \approx 5.90 \times 10^{-15}~\mathrm{A\,m/V^2}$.
}
  \label{fig:E3}
\end{figure}

\FloatBarrier  % Flushes figures here, but allows text to continue on the same page
\bibliography{paperRef}	

\end{document}